\documentclass[11pt,a4paper]{article} 
\usepackage{jheppub}

\usepackage{epsfig}
\usepackage{multicol}
\usepackage{amsmath,amsfonts,amssymb,amscd,amsthm,mathtools}

\usepackage{amsfonts}

\usepackage[english]{babel}
\usepackage[latin1]{inputenc}

\usepackage{epsfig}
\usepackage{rotate}

\usepackage{slashed}
\usepackage{graphicx}

\newcommand{\roughly}[1]{\mathrel{\raise.3ex\hbox{$#1$\kern-0.85em
\lower1ex\hbox{$\sim$}}}}

\def\be{\begin{equation}}
\def\beq\begin{equation}
\def\bea{\begin{eqnarray}}
\def\eea{\end{eqnarray}}

\def\beq{\begin{equation}}
\def\eeq{\end{equation}}
\def\beqa{\begin{eqnarray}}
\def\eeqa{\end{eqnarray}}

\def\dsl{\hbox{/\kern-.5300em$\partial$}}

\newcommand{\bmat}{\left(\begin{array}}
\newcommand{\emat}{\end{array}\right)}

\def\-{\hphantom{-}}

\def\s2{\frac{1}{2}}

\def\IF{\relax{\rm I\kern-.18em F}}
\def\II{\relax{\rm I\kern-.18em I}}
\def\IP{\relax{\rm I\kern-.18em P}}
\def\IC{\relax{\rm I\kern-.48em C}}
\def\IR{\relax{\rm I\kern-.18em R}}
\def\IK{\relax{\rm I\kern-.20em K}}
\def\IM{\relax{\rm I\kern-.25em M}}

\def\Dsl{\,\raise.15ex\hbox{/}\mkern-13.5mu D} 
\def \one{\relax{\rm 1\kern-.26em I}}

\usepackage{marginnote}

\newcommand\pair{e^+ e^-}

\newcommand\ee{e^+e^-}
\newcommand\g{\gamma}

\newcommand\pp{{A'}}

\newcommand\vdecay{A' \rightarrow e^+  e^-}
\newcommand\xdecay{A' \rightarrow e^+  e^-}

\newcommand\aee{A' \rightarrow \ee}
\newcommand\ainv{A' \rightarrow invisible}
\newcommand\ma{m_{A'}}
\newcommand\inv{\chi \bar{\chi}}

\title{Proposal for an Experiment\\
to Search for Light Dark Matter at the SPS}

\author{ 
S.~Andreas$^{a,b}$, S.V.~Donskov$^c$, P.~Crivelli$^d$, A.~Gardikiotis$^e$, S.N.~Gninenko$^f$\footnote{Contact person, Sergei.Gninenko@cern.ch}, \\ N.A.~Golubev$^f$, F.F.~Guber$^f$, A.P.~Ivashkin$^f$, M.M.~Kirsanov$^f$, N.V.~Krasnikov$^f$,\\
V.A.~Matveev$^{f, g}$, Yu.V.~Mikhailov$^c$, Yu.V.~Musienko$^f$, V.A.~Polyakov$^c$, A.~Ringwald$^a$, A.~Rubbia$^d$, V.D.~Samoylenko$^c$, Y.K.~Semertzidis$^h$, K.~Zioutas$^e$\\
\vskip1.cm
$^a${\it Deutsches Elektronen-Synchrotron DESY, 22607 Hamburg, Notkestrasse 85, Germany}\\
$^b${\it Institut d'Astrophysique de Paris IAP, 75014 Paris, France} \\
$^c$ {\it State Research Center of the Russian Federation, Institute for High Energy Physics, 142281 Protvino, Russia}\\
$^d${\it ETH Zurich, Institute for Particle Physics, CH-8093 Zurich, Switzerland}\\
$^e${\it Physics Department, University of Patras, Patras, Greece}\\ 
$^f${\it Institute for Nuclear Research, Moscow 117312, Russia}\\
$^g${\it Joint Institute for Nuclear Research, 141980 Dubna, Russia}\\
$^h${\it Center for Axion and Precision Physics, IBS, Physics Dept., KAIST, Daejeon, Republic of Korea}
\vskip6.cm
\begin{center}
\today
\end{center}
}

\abstract{\\
Several models of dark matter suggest the existence of dark sectors consisting of $SU(3)_C \times SU(2)_L \times U(1)_Y$ singlet fields. These sectors of particles do not interact with the ordinary matter directly but could couple to it via gravity. In addition to gravity, there might be another very weak interaction between the ordinary and dark matter mediated by $U'(1)$ gauge bosons $A'$ (dark photons) mixing with our photons. In a class of models the corresponding dark gauge bosons could be light and have the $\g - A'$ coupling strength laying in the experimentally accessible and theoretically interesting region. If such $A'$ mediators exist, their di-electron decays $\aee$ could be searched for in a light-shining-through-a-wall experiment looking for an excess of events with the two-shower signature generated by a single high energy electron in the detector. A proposal to perform such an experiment aiming to probe the still unexplored area of the mixing strength $10^{-5}\lesssim \epsilon \lesssim 10^{-3}$ and masses $M_{A'} \lesssim 100$ MeV by using 10-300 GeV electron beams from the CERN SPS is presented. The experiment can provide complementary coverage of the parameter space, which is intended to be probed by other searches. It has also a capability for a sensitive search for $A'$s decaying invisibly to dark-sector particles, such as dark matter, which could cover a significant part of the still allowed parameter space. The full running time of the proposed measurements is requested to be up to several months, and it could be taken at different SPS secondary beams. 
}

\begin{document}
\maketitle
\addcontentsline{toc}{section}{Table of Contents}
\newpage

\section*{Executive summary}
We propose an experiment dedicated to the sensitive search for the decays $\aee$ of massive dark photons ($A'$) into  $\ee$ pairs.  If $A'$s with the $\g - A'$ mixing strength in the range $10^{-5}\lesssim \epsilon \lesssim 10^{-3}$ and masses $M_{A'} \lesssim 100$ MeV exist, they could be observed through the $A'$ production in the reaction $e^- Z \rightarrow e^- Z A'$ of electrons scattering off nuclei, followed by the decay $\aee$. The experimental signature of this process - the two-shower energy deposition in the detector -  has never been experimentally tested before.

The new experiment could exploit one of the secondary beam  lines at the CERN SPS, which can provide electrons with an energy up to $\simeq 300$ GeV. The detector consists of a compact, specially designed  scintillator-tungsten  electromagnetic (e-m) calorimeter of a high longitudinal hermeticity, additionally protected agains the energy leak  by high efficiency veto counters. It is also equipped with the $A'$ decay volume, scintillating fiber  trackers, and beam defining scintillator counters and wire chambers,  which  provide information for tagging the incoming particles. 
 
Event candidates  with two e-m showers, which could originate from the $A'$ production and subsequent decay, are selected. The analysis based on the kinematic and topological shower properties is used to separate the signal from the background, dominated by the hadronic contamination in the beam. The feasibility study of the experimental setup shows that a sensitivity for the search of the $\aee$ decay mode  in branching fraction $Br(A') = \frac{\sigma(e^-Z\rightarrow e^-Z A')}{\sigma(e^-Z \rightarrow e^- Z \g)}$ at the level below a few parts in $10^{12}$  could be achieved. This would allow  to cover a significant fraction of the yet unexplored parameters space.  
 
The experiment has also a capability to search for  invisible decays  $\ainv$ with a high sensitivity. The feasibility study shows that a sensitivity for the search of the $\ainv$ decay mode in branching fraction $Br(A') = \frac{\sigma(e^-Z\rightarrow e^-Z A'), \ainv}{\sigma(e^-Z \rightarrow e^- Z \g)}$ at the level below a few parts in $10^{11}-10^{12}$ could be achieved. The intrinsic background due to the presence of low energy electrons in the beam can be  suppressed by using a tagging system, which is  based on the detection of synchrotron radiation of high energy electrons. The search would  also allow  to cover a significant fraction of the yet unexplored  parameters space for the $\ainv$ decay mode.

After testing the detector, that might commence in  2015, the experiment would be performed in two phases. In the first phase in 2015, the goal is to optimize the detector components and measure the dominant backgrounds from  the hadron (and possibly muon) contaminations in the electron beam. This could be done  by using any secondary beam line of the SPS that would provide enough intensity in the given energy range for the background measurements. In the second phase, 2015-2016,  the goal is to reach the previously mentioned  sensitivity  or better by exploiting a possible  upgrade of the detector, which might be necessary given the results of phase I. To reach this goal utilizing  a secondary SPS beam line that would provide enough electron intensity for the signal search is mandatory.   If an excess consistent with the signal hypothesis is observed, this would unambiguously indicate the presence of new physics.  

\newpage

\section{Introduction and Motivation}

Cosmological observations of galactic rotational curves~\cite{Borriello:2000rv} and the gravitational lensing~\cite{Hoekstra:2002nf,Metcalf:2003sz:Moustakas:2002iz} give strong evidence for the existence of dark matter (see e.g.~\cite{Bertone:2004pz} for a review). The challenge to explain these hints of the existence  of dark matter provides one of the strongest indications for the existence of new physics beyond the Standard Model (SM). The identification of the origin of dark matter (DM) is a problem of enormous importance for both particle physics and cosmology. At present, the most popular candidates for the thermal-produced DM  are the so-called WIMPs (weakly interacting massive particles), which are e.g.\ lightest supersymmetric particles, Kaluza-Klein particles in universal extra dimension models etc... However, despite of significant efforts the experiments, in particular at the LHC,  searching  for WIMPs lead so far to negative results, thus, pushing further WIMP searches into a very high-energy and/or high sensitivity frontiers, for a review see e.g.~\cite{Feng:2010gw} and references therein.

An additional natural ground for  understanding of the origin and properties  of dark matter is provided by a class of interesting theoretical models introducing the concept of ``dark'' (or hidden)  sectors consisting of $SU(3)_C \times SU(2)_L \times U(1)_Y$ singlet fields. These  sectors   of particles do not interact with the SM matter directly and  couple to it by gravity and possibly by other weak forces. It is worthwhile to note that, even in the SM, some fields of the matter are singlet under one or more of the colour and electroweak gauge groups. Thus, the idea  to include a further sector which transforms under the new but not under the familiar gauge symmetries is not particularly exotic from a theoretical viewpoint. The sensitivity of experiments searching for the new singlet particles depends in detail on their couplings and mass scale, for instance,  if the mass scale of a dark sector is too high, it is experimentally unobservable and indeed is hidden.

Then, one could ask a natural  question: could the important sensitive searches for the dark sectors be performed at lower energy and high intensity  frontier? The answer for this question is definitely positive. For example, there is a class of models with at least one additional U(1) gauge factor where the corresponding hidden gauge boson could be light, or even massless~\cite{Okun:1982xi,Holdom:1985ag,Ahlers:2007rd:Ahlers:2007qf:Jaeckel:2007ch:Caspers:2009cj:Ohta:2009un:Goodsell:2009xc:Jaeckel:2009wm:Mirizzi:2009iz,Gninenko:2008pz}, for a recent review see~\cite{Jaeckel:2010ni,Hewett:2012ns,Essig:2013lka}. The interaction between SM and dark matter may be  transmitted by a new abelian $U'(1)$ gauge  bosons $A'$ (or dark photons for short) mixing with ordinary photons, see e.g.~\cite{Pospelov:2007mp,Chun:2010ve,Mambrini:2011dw,Andreas:2011in,Hooper:2012cw,Davoudiasl:2013jma,Davoudiasl:2013aya,Jaeckel:2013ija}. The original idea was  first discussed by  Okun in his  paraphoton model~\cite{Okun:1982xi}, see also~\cite{Holdom:1985ag}. For the massless case, e.g.\ in the mirror dark matter models, the portal to our world through photon-mirror photon mixing  leads to orthopositronium (oPs) to mirror orthopositronium oscillations, the experimental signature of which is the apparently invisible decay of oPs, for review, references  and more detail discussions see~\cite{ar,Felcini:2004yg,Gninenko:2002jn,Gninenko:2006sz,Crivelli:2010bk}. 

Experimental bounds on the sub-eV and sub-keV  dark photons can be obtained  from searches for the fifth force~\cite{Okun:1982xi,Williams:1971ms,Bartlett:1988yy}, from experiments on  photon regeneration~\cite{phreg,VanBibber:1987rq,Sikivie:1983ip,Raffelt:1987im,vanBibber:1988ge,Redondo:2010dp,Betz:2013dza}, and from stellar cooling considerations \cite{seva1,seva2}. For example, it has been noticed that helioscopes searching for solar axions   are sensitive to the keV part of the solar spectrum of hidden photons and the CAST results~\cite{Zioutas:2004hi,Andriamonje:2007ew} have been translated into limits on the $\g - \pp$ mixing parameter~\cite{Redondo:2008aa,Redondo:2012ky,Gninenko:2008pz,Troitsky:2011rx}. Stringent bounds on models with additional $A'$ particles  at a low energy scale could be obtained from astrophysical considerations~\cite{Blinnikov:1990ui,Davidson:1993sj,Davidson:2000hf}. In some models these astrophysical constraints can be relaxed or evaded, see e.g.~\cite{Masso:2006gc}. New tests on the  existence of sub-eV $A'$s  at new experimental facilities, such, for example, as SHIPS~\cite{Schwarz:2011gu}, ALPS-II~\cite{Bahre:2013ywa} or IAXO~\cite{Irastorza:2012qf} are in preparation.

The $A'$s in the sub-GeV mass range, see e.g.~\cite{Pospelov:2007mp,Batell:2009di,Reece:2009un,Williams:2011qb,Davoudiasl:2013jma,Davoudiasl:2013aya,Andreas:2011in} can be probed  through the searches for $A'\rightarrow \ee$ decays in beam dump experiments~\cite{Bjorken:2009mm,Andreas:2012mt,Konaka:1986cb,Riordan:1987aw,Bjorken:1988as,Davier:1989wz,Bross:1989mp,Blumlein:2011mv,Merkel:2011ze,Abrahamyan:2011gv,Blumlein:2013cua,Beranek:2013nqa}, or through the rare particle decays, see e.g.~\cite{Archilli:2011zc,Li:2009wz,Aubert:2009cp,Gninenko:2013sr,MeijerDrees:1992kd,Adlarson:2013eza,Agakishiev:2013fwl}. For example, if the $A'$ mass is below the mass of $\pi^0$, it can be effectively searched for in the decays $\pi^0 \rightarrow \gamma A'$, with the  subsequent decay of $A'$ into an $\ee$ pair. Recently, stringent constraints on the mixing $\epsilon$ in sub-GeV mass range  have been derived  from a search of this  decay mode with existing data of neutrino experiments~\cite{Gninenko:2011uv,Gninenko:1998pm,Altegoer:1998qta} and from SN1987A cooling~\cite{Dent:2012mx}. In a class of models, the $A'$  may have mass $\ma \lesssim 100$ MeV and  $\gamma-A'$ mixing strength as large as $\epsilon \simeq 10^{-5} - 10^{-3}$, which is in the experimentally accessible and theoretically interesting region,~\cite{nah}.  This makes further searches for dark mediators interesting and attractive, for a recent review see~\cite{Jaeckel:2010ni, Hewett:2012ns, Essig:2013lka, Jaeckel:2013ija}, and references therein.

The main goal of this proposal is to investigate still  unexplored region of  mixing strength $10^{-5}\lesssim \epsilon \lesssim 10^{-3}$ and $A'$ masses $M_{A'}\lesssim 100$ MeV with  a light-shining-through-a-wall (LSTW) type experiment~\cite{Jaeckel:2010ni} using a high energy secondary  electron beam for the CERN SPS. If such $A'$s exist, they would be short-lived particles which decay  rapidly  into $\ee$ pairs with a lifetime in the range $10^{-14}\lesssim  \tau_{A'} \lesssim 10^{-10}$ s. We show that such decays could be observed by looking for events with the exotic signature - two isolated showers produced by a single electron in the detector. If, indeed, an excess of such events is observed, this would be a strong evidence for the existence of new physics  beyond the SM.

Compared to  beam-dump experiments searching for relatively long-lived $A'$s, the advantage of the proposed one is that its sensitivity is roughly proportional to the mixing squared, $\epsilon^2$, associated with the $A'$ primary production process (for a short lived particle the decay probability  inside the decaying volume is close to 1). For the long-lived $A'$ case, the sensitivity of the search  is proportional to $\epsilon^4$ - one $\epsilon^2$ came from the $A'$ production, and another $\epsilon^2$ is from their decays. Another advantage of the project is that the expected background level can be precisely determined from the direct measurements with pion and muon beams in the same setup.  Below, we present a conceptual detector scheme that would exploit one of the existing secondary  electron  beam, a scintillator-tungsten electromagnetic calorimeters, veto counters, and scintillator fiber tracker.

The proposed LSTW type experiment is not a new one for CERN. Historically, one of the first experiment of this type was performed at CERN in 2000 by the NOMAD collaboration. The search for light (pseudo)scalars ($a$) with the coupling to two photons was performed at high energies by using the NOMAD neutrino detector~\cite{Astier:2000gx,Gninenko:2000ds}. If $a$s exist, one expects a flux of such high energy particles in the SPS neutrino beam because both scalar and pseudoscalar $a$s could be produced in the forward direction through the Primakoff effect in interactions of high energy photons, generated by 450~GeV protons from the SPS  in the neutrino target, with virtual photons from the magnetic field of the  WANF horn. If $a$ is a relatively long-lived particle, it would penetrate the downstream shield without interaction and would be observed in the NOMAD detector via the inverse Primakoff effect, namely the interaction of (pseudo)scalars with virtual photons from the field of the NOMAD dipole magnet. The experimental signature of the $a-\g$ conversion is a  high-energy photon resulting in a single isolated electromagnetic shower in the NOMAD electromagnetic calorimeter. Later on, the new limits for  dark photons in the mass range $\lesssim 1$~eV, were set from results of the  CAST experiment at CERN~\cite{Redondo:2008aa, Redondo:2012ky,  Gninenko:2008pz}.  Recently, future potential for dark  photon physics has been discussed by the IAXO collaboration~\cite{Irastorza:2012qf}.

Finally, let us note that in addition to dark photon models, many extensions of the Standard Model (SM) such as GUTs~\cite{Langacker:1980js}, super-symmetric~\cite{Weinberg:1981wj:Fayet:1980rr}, super-string models~\cite{Ellis:1985yc,Cicoli:2011yh} and models including a new long-range interaction, i.e.\ the fifth force~\cite{Carlson:1986cu}, predict an extra, U$'$(1) factor and therefore the existence of a new gauge boson $X$ corresponding to this new group. The predictions for the  mass of the  $X$ boson are not very firm and it could be light enough ($M_{X}\ll M_{Z}$) to be searched for  at low energies. For instance, if the mass $M_X$ is  of the order of the pion mass, an effective search could be conducted for this new vector boson in the radiative decays of neutral  pseudoscalar mesons $P\rightarrow\gamma X$, where $P = \pi^{0},\eta$, or $\eta^{\prime}$, because  the decay rate of  $P\rightarrow\gamma\;+$ $\it any~new~particles~with~spin~0~or~\frac{1}{2}$ proves to be negligibly small~\cite{Dobroliubov:1987cb:Dobroliubov:1988pe}. Hence, a positive result in the direct search for these decays could be interpreted unambiguously as the discovery of a new light spin 1 particle, in contrast with other experiments searching for light weakly interacting particles in rare K, $\pi$ or $\mu$ decays~\cite{Dobroliubov:1987cb:Dobroliubov:1988pe,Dobroliubov:1990ye:Dobroliubov:1991kg,Gninenko:2001hx}. Such light $X$s coupled, e.g.\ to leptons and quarks could be searched for in an analogous  LSTW experiment with a high energy pions, see e.g.~\cite{Gninenko:2011uv}.
 
The rest of the document is organized in the following way. The theoretical considerations of the $A'$ production and decay are presented in Sec.~\ref{sec:Theory}. The experimental setup, method of search, and requirements to the beam as well as background sources and the expected sensitivity are described in Sec.~\ref{sec:ExpVisible}. Section~\ref{sec:ExpInvisible} contains a discussion of the search for an $A'$ which decays invisibly into two dark matter particles as well as the corresponding backgrounds and the estimated sensitivity. Section~\ref{sec:Conclusions} contains concluding remarks.

\section{Theoretical considerations} \label{sec:Theory}
The interaction between $\g$'s and $A'$'s is  given by the kinetic mixing~\cite{Okun:1982xi, Holdom:1985ag, Jaeckel:2010ni}  
\begin{equation}
 L_{int}= -\frac{1}{2}\epsilon F_{\mu\nu}A'^{\mu\nu} ,
\label{mixing}
\end{equation}
where  $F^{\mu\nu}$, $A'^{\mu\nu}$ are the ordinary  and the  dark photon  fields, respectively, and parameter~$\epsilon$ is their mixing strength. The kinetic mixing of Eq.~\eqref{mixing} can be diagonalized resulting for massive $A'$ in a nondiagonal mass term and $\gamma - A'$ mixing. Therefore, {\it any source of photons} could produce a kinematically permitted massive $A'$ state   according to the  mixings. Then, depending on the $A'$ mass, photons may  oscillate into dark photons -- similarly to  oscillations of neutrinos -- or,  the $A'$'s could  decay, e.g.\ into $\ee$ pairs. 
\begin{figure}[tbh!]
\begin{center}
\includegraphics[width=0.6\textwidth]{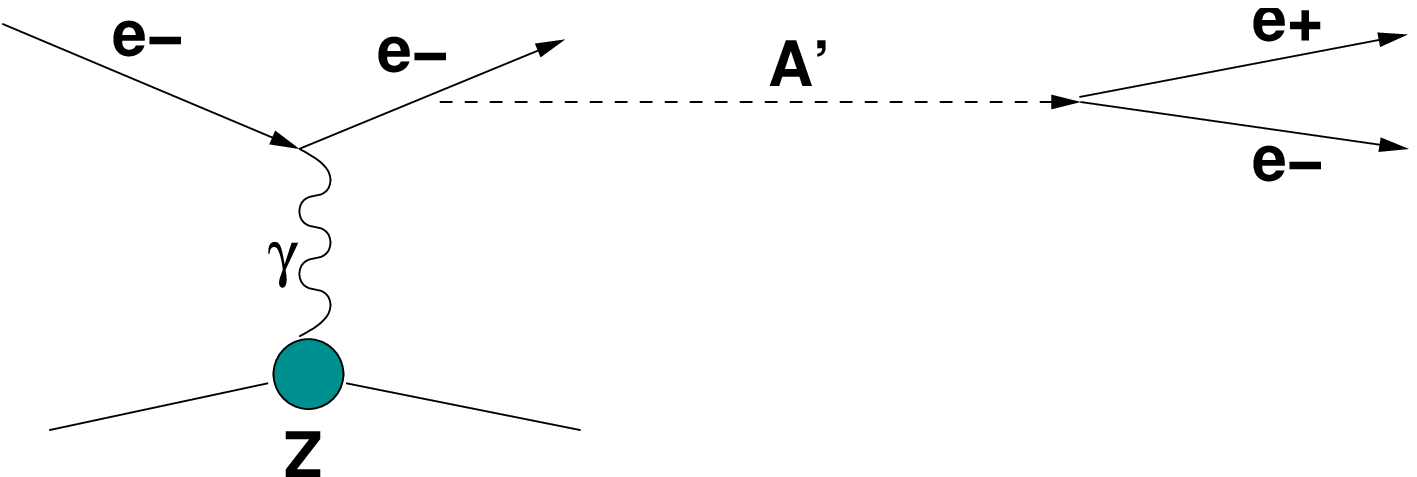}
\caption{Diagram illustrating the massive $A'$ production in the  reaction $e^- Z \rightarrow e^- Z A'$ of electrons scattering off a nuclei (A,Z) with the subsequent $A'$ decay into an $\ee $ pair.\label{diagr}}
\end{center}
\end{figure}
The diagram for the $A'$ production in the reaction 
\begin{equation}
 e^- Z \rightarrow e^- Z A',~\aee 
\label{reaction}
\end{equation}
is shown in  Fig.~\ref{diagr}. The total number of $A'$s produced by $n_e$ electrons in a target with thickness $t\gg X_0$  is~\cite{Bjorken:2009mm}: 
\begin{equation}
n_{A'} \sim n_e C \frac{ \epsilon^2 m_e^2}{M_{A'}^2} \, ,
\label{natot}
\end{equation}
where the parameter $C\simeq 10$ is only logarithmically dependent on the choice of target nucleus, and $m_e$ is the electron mass and $M_{A'}$ the $A'$ mass, for recent works on heavy particles production through photon exchange with a nucleus,  see also \cite{Masip:2012ke,Radionov:2013mca}. In~\cite{Andreas:2012mt,Andreas:2013xxa} it is argued, that the parameter $C$ is actually $C\simeq 5$. One can see that compared to bremsstrahlung rate, the $A'$ production is suppressed by a factor $\simeq \epsilon^2 m_e^2/M_{A'}^2$. Therefore, for the parameter space region of our interest, it is  expected to occur with the rate $\lesssim 10^{-13}-10^{-9}$ with respect to the  ordinary photon production rate. 
 The $A'$ energy spectrum is~\cite{Bjorken:2009mm} 
\begin{equation}
\frac{d n_{A'}}{dE_{A'}} \sim k\cdot x \bigl(1+\frac{x^2}{3(1-x)}\bigr) \, ,
\end{equation}
where $k$ is a constant, and $x=E_{A'}/E_0$. 

The $A'$ is emitted  with respect to electron beam axis dominantly  at an angle $\Theta_{A'} \lesssim \Theta_{\ee} \simeq \ma/E_{A'}$, which is  is typically smaller than the opening angle of the $\aee$ decay products $\Theta_{\ee}$. The approximation of $A'$ emission collinear with the beam axis  is justified in many calculations~\cite{Bjorken:2009mm}.

The corresponding $\aee$ decay rate is given by:
\begin{equation}
\Gamma (\vdecay) = \frac{\alpha}{3} \epsilon^2 M_{A'} \sqrt{1-\frac{4m_e^2}{M_{A'}^2}} \Bigl( 1+ \frac{2m_e^2}{M_{A'}^2}\Bigr) \, .
\label{rate}
\end{equation}  
It is assumed that this decay mode is dominant and the branching ratio $\frac{\Gamma(\aee)}{\Gamma_{tot}}\simeq 1$.
\begin{figure}[tbh!]
\begin{center}
\includegraphics[width=0.6\textwidth]{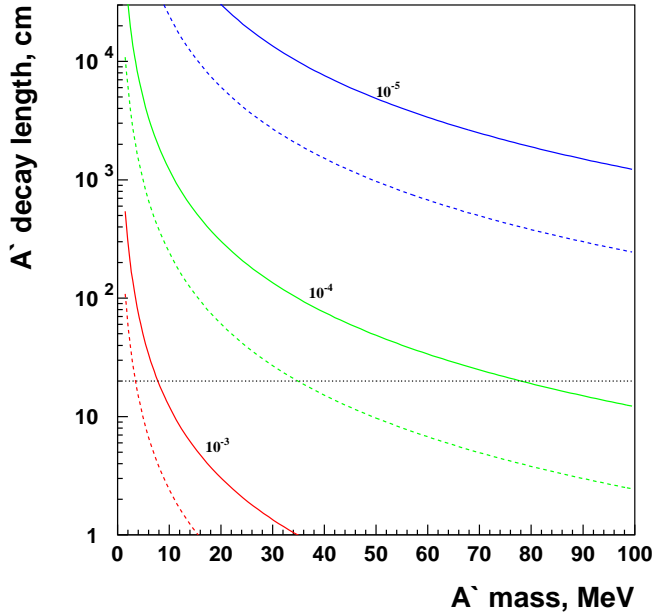}
\caption{The $A'$ decay length as a function of its mass calculated for different $\g-A'$ mixing values indicated near the curves and for the $A'$ energy of 150 (solid) and 30 (dashed) GeV. The horizontal line (dotted) indicates the approximate length of the designed calorimeter ECAL1, $\simeq 200$~mm.\label{decay}}
\end{center}
\end{figure}


\section{The experiment to search for the decay \texorpdfstring{$\aee$}{A' -> e+e-}} \label{sec:ExpVisible}

The process of the dark photon production and subsequent decay is a rare  event as pointed out above. Hence, its observation presents a challenge for the detector design and performance. 

\subsection{The Setup}

\begin{figure}
\includegraphics[width=0.9\textwidth]{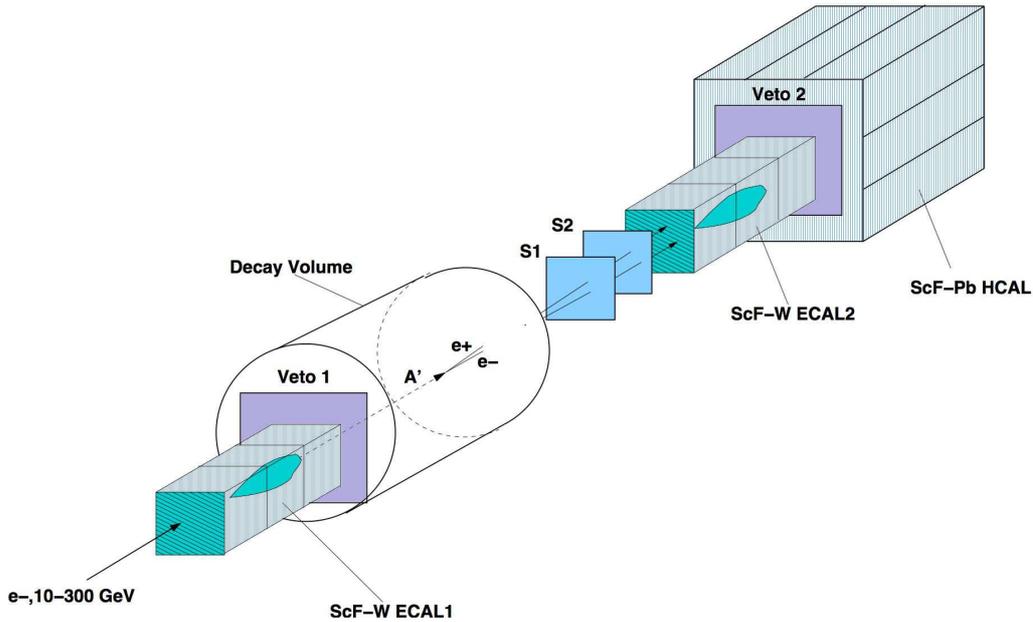}
\caption{Schematic illustration of the setup to search for dark photons in a light-shining-through-a-wall type  experiment at high energies. The incident electron energy absorption in the calorimeter ECAL1 is accompanied by the emission  of  bremsstrahlung $A'$s in the reaction  $eZ\rightarrow eZ A'$ of electrons scattering on nuclei, due to  the $\g - A'$ mixing. The part of the primary beam energy is deposited in the ECAL1, while  the  rest of the total energy is  transmitted by the $A'$ through the ``ECAL1 wall''. The $A'$ penetrates the ECAL1 and veto V1  without interactions and  decays in flight in the decay volume DV into a narrow $\ee$ pair, which  generates the second electromagnetic shower in the ECAL2 resulting in the two-shower  signature in the detector. The sum of energies deposited in the ECAL1+ECAL2 is equal to the primary beam energy. This detector, being additionally equipped with the  hadronic calorimeter (HCAL) to enhance its longitudinal hermeticity, can also be  used to search for the invisible decay $\ainv$ of dark photons into the lighter dark matter particles $\chi$, see Sec.~\ref{sec:ExpInvisible}.\label{setup}}
\end{figure}

The experimental setup specifically designed to search for the  $A'\rightarrow \ee$ decays is schematically shown in Fig.~\ref{setup}. The experiment requires a clean high energy  $e^-$ beam, with impurities below the one percent level. The primary proton energy of 400 GeV from the SPS enables secondary electron beams in the energy range from 10 to 300 GeV with typical intensities ranging from $10^7$ down to $10^5$ electrons per SPS spill~\cite{sps}. The admixture of other charged particles in the beam (beam purity) is below $10^{-2}$. The detector shown in Fig.~\ref{setup} is equipped with a  high density, compact electromagnetic (e-m) calorimeter ECAL1 to detect $e^-$ primary interactions. The counters V1 and V2 serve as high efficiency veto while the two scintillating fiber counters (or proportional chambers) S1, S2 and an electromagnetic  calorimeter ECAL2 located at the downstream end of the $A'$ decay volume DV will detect $\ee$ pairs from $\aee$ decays in flight. For searches at low energies the DV could be replaced by a Cherenkov counter to enhance the tagging efficiency of the decay electrons.

The method of the search is the following \cite{Gninenko:2013rka}. The $A'$s are produced through the mixing with bremsstrahlung photons from the electrons scattering off nuclei in the ECAL1. The reaction \eqref{reaction} typically occurs within the first few radiation length ($X_0$) of the detector. The bremsstrahlung $A'$ then penetrates the rest of the ECAL1 and the veto counter V1 without interactions,  and decays in flight into an $\ee$ pair in the decay volume DV. A fraction ($f$) of the primary beam energy $E_1 = f E_0$  is deposited in the ECAL1. The ECAL1's downstream part is served  as a dump to absorb completely the e-m shower tail. For the radiation length $\lesssim$ 1 cm, and the total thickness of the ECAL1 $\simeq 30~X_0$ (rad.\ lengths) the energy leak  from the ECAL1 into the V1 is  negligibly small. The remained part of the primary electron energy $E_2 = (1-f)E_0$ is transmitted trough the ``ECAL1 wall'' by the $A'$, and deposited in the second downstream calorimeter ECAL2  via the $A'$ decay in flight in the DV, as shown in Fig.~\ref{setup}.  At high $A'$ energies $E_{A'}\gtrsim 30$ GeV,  the opening angle  $\Theta_{\ee} \simeq M_{A'}/E_{A'}$ of the decay $\ee$ pair is too small to be resolved in two e-m showers in the ECAL2, so the pairs are mostly  detected as a single electromagnetic  shower. At distances larger than $\simeq$ 5 m from the ECAL1, the distance between the hits is  $\gtrsim 5$ mm, so the $\ee$ pair  can be resolved in two separated tracks in the S1 and S2.  

\begin{figure}
\begin{center}
\includegraphics[width=0.7\textwidth]{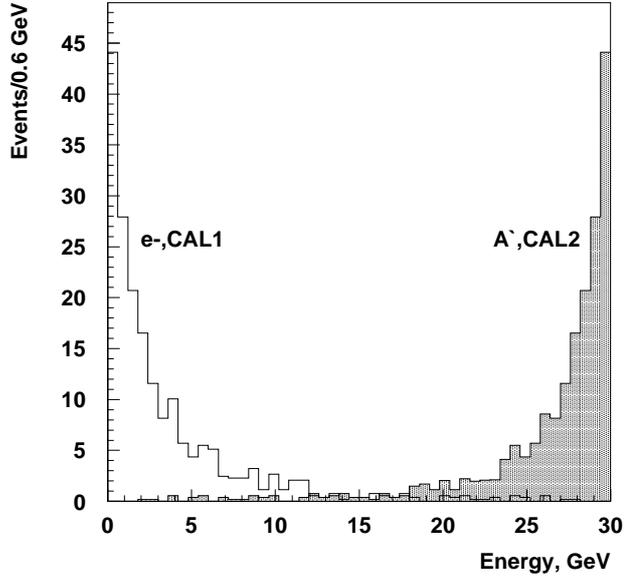}
\caption{Expected distributions of energy deposition for selected events: i) in the ECAL1, and ii) in the ECAL2 (shaded) from the bremsstrahlung $\aee $ decays in flight in the DV region. The spectra are calculated for the 10 MeV $A'$s produced by 30 GeV e$^-$s in the ECAL1 with momentum pointing towards the ECAL2  fiducial area and the mixing strength $\epsilon = 3\cdot 10^{-4}$. For this mixing value most of $A'$s decay outside of the ECAL1 in the DV. The distributions are normalized to a common maximum.\label{energy}}
\end{center}
\end{figure}

The occurrence of $\xdecay$ decays produced in $e^- Z $ interactions would appear as an excess of events with two e-m-like showers in the detector, Fig.~\ref{setup}, above those expected from the background sources. The signal candidate events have the signature:  
\begin{equation}
S_{A'} = {\rm ECAL1 \times \overline{V1} \times S1 \times S2 \times ECAL2 \times \overline{V2}},
\label{sign}
\end{equation}
and should satisfy the following selection criteria: 
\begin{itemize}
\item The starting point of (e-m) showers in the ECAL1 and ECAL2 should be localized  within a few first $X_0$s.  
\item The lateral and longitudinal shapes of both showers in the ECAL1 and ECAL2 are consistent with an electromagnetic one. The fraction of the total  energy deposition in the ECAL1 is $f\lesssim 0.1$, while in the ECAL2 it is $(1-f)\gtrsim 0.9$ (see energy spectra in Fig.~\ref{energy}, and discussion below).
\item No energy deposition in  the V1 and V2.
\item The signal (number of photoelectrons) in the decay counters S1 and S2 is consistent with the one expected  from two minimum ionizing particle (mip) tracks. At low beam energies, $E_0\lesssim 30$ GeV,  two isolated hits in each counter are requested. 
\item the sum of energies deposited in the ECAL1+ECAL2 is equal to the primary energy,  $E_1 +E_2 = E_0$.
\end{itemize}

In Fig.~\ref{energy} an example of the expected  distributions of energy deposition in the ECAL1 and ECAL2 for selected events are shown for the initial $e^-$ energy of 30 GeV. The spectra are calculated  for the mixing strength $\epsilon = 3\cdot10^{-4}$ and correspond to the case when the $A'$ decay pass length $L_{A'}$ is in the range $L' < L_{A'}< L$, where $L'$ is the length of the ECAL1 and $L$ is the distance between the ${A'}$ production vertex and the ECAL2. In this case  most of $A'$s decay outside of the ECAL1 in the DV. One can see, that the $A'$ bremsstrahlung distribution is peaked at maximal beam energy.  The measurement of the electron energy and the background level could be  deteriorated by the presence of passive material in the detector. Therefore, the passive material budget must be minimized. The detector, being additionally equipped with the  hadronic calorimeter (HCAL) to enhance its longitudinal hermeticity as shown in Fig.~\ref{setup}, can also be  used to search for the invisible decay $\ainv$ of dark photons into the lighter dark matter particles $\chi$, see Sec.~\ref{sec:ExpVisible} for discussions.

\subsubsection{The SPS H4 secondary beam line} \label{sec:SPSbeamline}

\begin{figure}[tbh!]
\begin{center}
\includegraphics[width=0.7\textwidth]{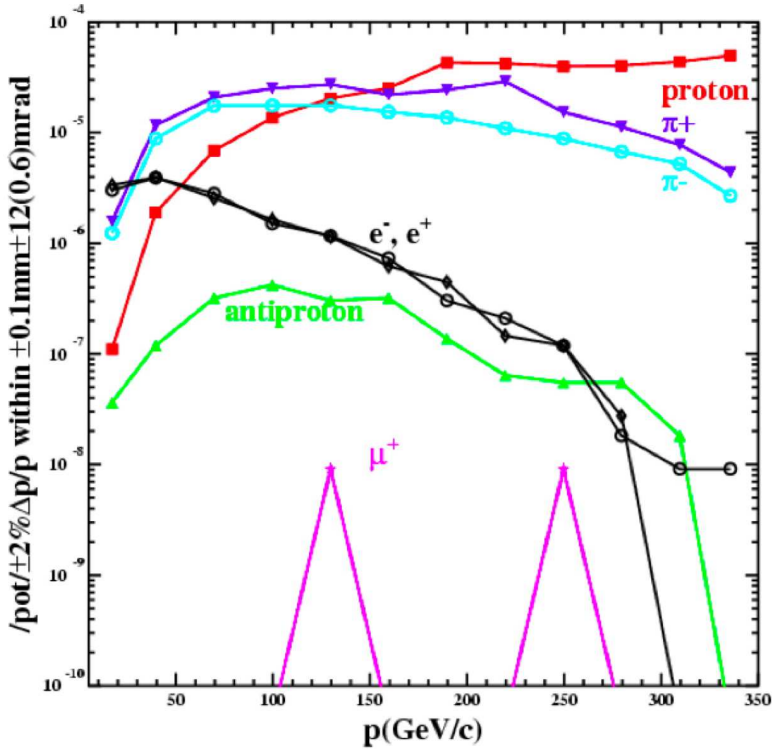}
\caption{Production rate of electrons (positrons) as a function of their energy  at the H4 secondary beam line from the primary T2 target~\cite{sps}.\label{eflux}}
\end{center}
\end{figure}

The experiment could employ, e.g.\ the CERN SPS  H4 $e^-$ beam, which is produced in the target T2 of the CERN SPS and transported to the detector in an evacuated beamline tuned to a freely ajustable  beam momentum from 10 up to 300 GeV/c. The typical maximal beam  intensity at $\simeq$ 30-50 GeV, is of the order of $ 5\times 10^6~e^-$ for one typical SPS spill with $10^{12}$ protons on target, see Fig.~\ref{eflux}, \cite{sps}. Note, that a typical SPS cycle for Fixed Target (FT) operation lasts 14.8 s, including 4.8 s spill duration. The maximal number of FT cycles is 4 per minute, however, this number can vary from 1 to 2 per minute.    

To provide as maximal as possible coverage of still unexplored area of the mixing strength $10^{-5}\lesssim \epsilon \lesssim 10^{-3}$ and masses $M_{A'} \lesssim 100$ MeV, and also to have realistic beam exposure time, we plan to take  measurements with a beam of 30-50 GeV with the total number of accumulated electrons on the ECAL1 $N^e_{tot}\gtrsim 10^{12}-10^{13}~ e^-$'s.  Reaching this goal requires an  average beam intensity of $\gtrsim 5\times 10^6~ e^-$ per SPS spill. Because, there are no special requirements for the small beam size  at the entrance to the detector, which could be  within a few cm$^2$, the beam intensity can be increased by a factor 2 by tuning the beam line optics and collimators. However, it is assumed that the contamination of particles, others than electrons is still within a few times $10^{-2}$. Thus, we can assume that for an optimistic scenario, the total  number of electrons accumulated during one month of data taking is $N^e_{tot}\simeq 2\times 10^{12}$. In a less optimistic case, this number could lay in the range $ 3 \times 10^{11} \lesssim N^e_{tot} \lesssim 2 \times 10^{12}$. Therefore, to accumulate $N^e_{tot} \gtrsim 10^{12}$ electrons, the data taking period of at least 3 months is requested.

The suppression of any possible background should be at a level of $10^{-12}$ or below. The advantage to use the H4 beam is that at high energies ($\gtrsim$ 30 GeV) the beam is  very clean, the contamination of $\pi$s in electron beam is expected to be well below 1\%. In the analysis  presented below, no special treatment was applied to the simulated data to eliminate an eventual pion contamination. The assumed further beam purity is $\simeq 10^{-2}$.

A two-stage approach is envisaged for the experiment, incorporating an initial experimental test phase in 2014-2015, followed by the  main-goal period of the experiment to reach sensitivity of $Br(A') \lesssim 10^{-12}$ in 2015-2016. Upstream of the detector shown in Fig.~\ref{setup}, a beam trigger counter telescope is  installed. It consists of several scintillation counters  (not shown in Fig.~\ref{setup}). Two MWPC chambers with X,Y read-out, situated at 5 m from each other, are used to define the beam impact point into the calorimeter ECAL1. The beam electrons are focused onto the center of the front area of the ECAL1. In order to reduce the noise, we plan to  apply a cut on all X and Y beam chamber profiles. In addition, to guarantee the direction of the beam to be parallel to the module axis, we require that the difference in X and Y measured by each of these chambers is smaller than 1 mm ($\simeq 0.2$ mrad).

\subsubsection{Veto and S1, S2 counters}  

The decay volume is followed by scintillating fiber hodoscopes counters  S1 and S2 and scintillating tiles coupled to silicon photomultipliers (SiPM). The main task of the system is to measure precisely the time of arrival of particles in order to allow the matching with hits detected in the S1 and S2 and to reject pile-up events. As illustrated in Fig.~\ref{setup} the S1, S2 system is composed of scintillating fibers (SciF) hodoscopes, rectangular in shape, with dimensions $\simeq$ 100 x100 $mm^2$. The electron tracks will be measured by scintillating fiber hodoscopes, arranged  upstream and downstream of the central SciF hodoscope. The thickness of the hodoscope is of order 1 mm. This is also the thickness that will be seen by the outgoing electron-positron pair produced in $A'$ decays, and has to be kept as low as possible, compatible with the proposed performance of this sub-detector. 

The scintillating light produced in the fibers will be detected with arrays of silicon photomultipliers (SiPM) at both fiber ends. The choice of SiPMs as a photodetection device is based on the fact that they are very compact detectors that can be operated (in high magnetic fields) with high gain ($\simeq10^6$) and at high counting rates ($>$ 1 MHz). Typical dimensions of such SiPM arrays available from Hamamatsu (or KETEK~\cite{musa}) are $\simeq$ 8 mm wide and 1 mm high, with $50 \times 50~\mu m^2$ or $100 \times 100~\mu m^2$ pixels. The pixels are arranged in columns, corresponding to an effective readout pitch of 250 microns. The photodetectors would be directly coupled to the SciF arrays to maximize the light collection efficiency. To readout the fiber hodoscope at each end a total of 20 such photodetectors will be required corresponding to about 400 readout channels.

For the detector we aim at a time resolution of 300 ps. Time resolutions of about 300 ps have already been achieved with ScF hodoscopes with single-sided readout using multianode PMTs~\cite{musa}. The veto counters are assumed to be 1-2 cm thick,  plastic scintillator counters with a high light yield of $\simeq 10^2$ photoelectrons per 1 MeV of deposited energy. The typical veto's inefficiency for the mip detection is, conservatively, $\lesssim 10^{-4}$. Each  of the decay counters S1 and S2  consists of two layers of scintillating fiber strips, arranged respectively in the X and Y direction. Each strip consists of about 100 fibers of 1 mm square. The number of photoelectrons produced by a mip crossing the strip is $\simeq$ 20 ph.e. In the design and construction of this detector it will be very important to maximize the photon detection efficiency of the photodetector in order to maximize the veto efficiency  and time resolution.

\subsubsection{The tungsten scintillator  calorimeters}

The choice of the calorimeter type should satisfy the following criteria: 
\begin{itemize} 
\item One of the main requirement for the sensitive search for $A'$s in the still unexplored parameter space, is to achieve a highly compact design, having a small Moliere radius and short radiation length. The total length of the detector should be $\lesssim 30 $ cm. This implies having the greatest amount of absorber possible, consistent with obtaining the required energy resolution.

\item The energy resolution should be $\Delta E/E \simeq 15\%/\sqrt{E}$.

\item It should be possible to measure the lateral and longitudinal shower shape.

\item  The $e/\pi$ rejection should be $\lesssim 10^{-3}$.

\item Timing properties should allow high speed data accumulation.

\item The radiation hardness must be better then 1000 Gy.

\end{itemize}
The energy resolution of the ECAL1 and ECAL2 calorimeters as a function of the beam energy is taken to be $\frac{\sigma}{E} = \frac{15\%}{\sqrt{E}} \oplus 3\% \oplus \frac{142~MeV}{E}$~\cite{Woody:2011bja}. 
 
\begin{figure}
\begin{center}
\includegraphics[width=0.8\textwidth]{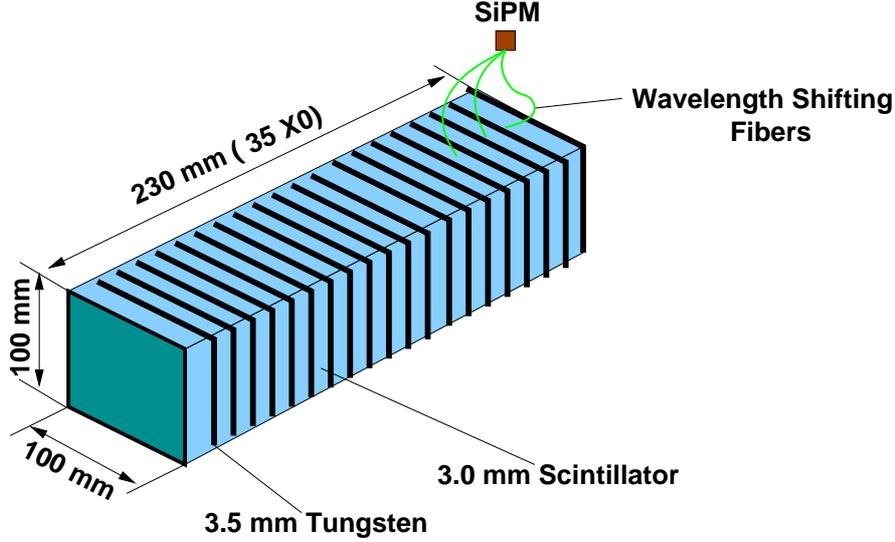}
\caption{Schematic illustration of a scintillator-fiber-tungsten module consisting of a stack of tungsten and scintillator plates of the size 3.5 mm and 3 mm, respectively. Wavelength shifting fibers pass laterally  through the plates and are read out at the side of the module  with either SiPMs or APDs photodetector. The similar module with a lateral 2x2 (or 3x3) cells segmentation is also under design.\label{ecalmodule}}
\end{center}
\end{figure}

To fulfill these design requirements, we are considering a scintillator - tungsten sandwich configuration, as shown in Fig.~\ref{ecalmodule}.
It consists of a standard sandwich arrangement of alternating W absorber and scintillator plates read out with wavelength shifting fibers running laterally through each scintillator plate. For example, the NA-61 hadronic calorimeter uses this type of design, but with lead absorber plates rather than tungsten~\cite{Golubeva:2009zza:Ivashkin:2012fd}. Our ECAL1 module design would be similar except it would have a more fine granularity, a higher density and make a more compact calorimeter with a smaller overall module size (roughly 10 square Moliere radius at the front). This design  would have the advantage of readout lateral and longitudinal shower profile, by utilizing  fibers from each plate (or a group of adjacent plates) to read out and also having better light collection. The ECAL1 is $\simeq 100\times 100$ mm$^2$ in cross section and 230 mm ($\simeq$35 $X_0$) long, see Fig.~\ref{ecalmodule}. Timing and energy deposition information from each plate can be digitized for each event. The processing of the counter signals is described in Sec.~\ref{sec:DAQ}. The possibility of using as the ECAL1 and ECAL2 the hodoscope arrays of the lead tungstate (PWO) heavy crystal counters ($X_0 \simeq 0.89$ cm), each of the size $10\times 10 \times 300$~mm$^3$, is also under consideration.

To evaluate the basic performance characteristics of this design we have carried out a Monte Carlo study by using GEANT4~\cite{Agostinelli:2002hh:Allison:2006ve}. For the calorimeter design, the energy resolution requirements are quite stringent and are in the range of a few \% for the energy region 30-100 GeV. 

\begin{figure}
\begin{center}
\includegraphics[width=0.7\textwidth]{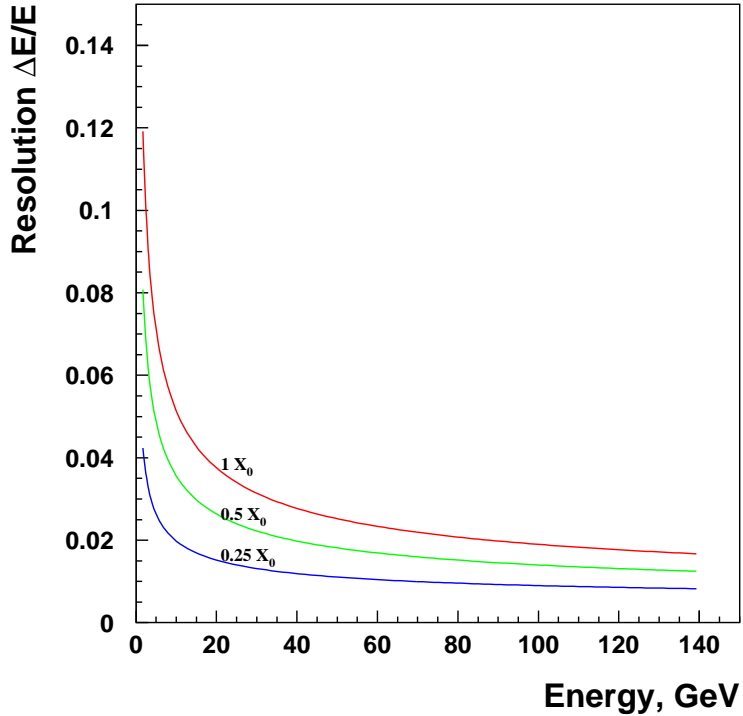}
\vspace{0.cm}\caption{Simulated energy resolution (FWHM) as a function of the incident energy for a  calorimeter module configuration shown in Fig.~\ref{module} for different absorber plate thicknesses, indicated near the curves. The scintillator plate thickness was kept constant at 3.0 mm for each configuration. Only contributions from sampling fluctuations and energy leakage are included.\label{resol}}
\end{center}
\end{figure}

We studied the ECAL1 energy resolution for various tungsten plate thicknesses keeping the scintillator thickness constant at 3.0 mm. Fig.~\ref{resol} gives the results of these simulations. The curves were fit to a parametrization $\Delta E/E = a/\sqrt{E} + b$ and the results of the fits for the selected W plate thickness of 3.5 mm is $a=0.15$ and $b=0.004$. It shows that an energy resolution $\simeq 15\%/\sqrt{E}$ can be achieved with the selected sampling. Note, that only sampling fluctuations and leakage were included in this simulation, therefore the photo-statistics contribution has to be kept small compared to this value. 

\begin{figure}[tbh!]
\begin{center}
\hspace{-0.cm}\includegraphics[width=.7\textwidth]{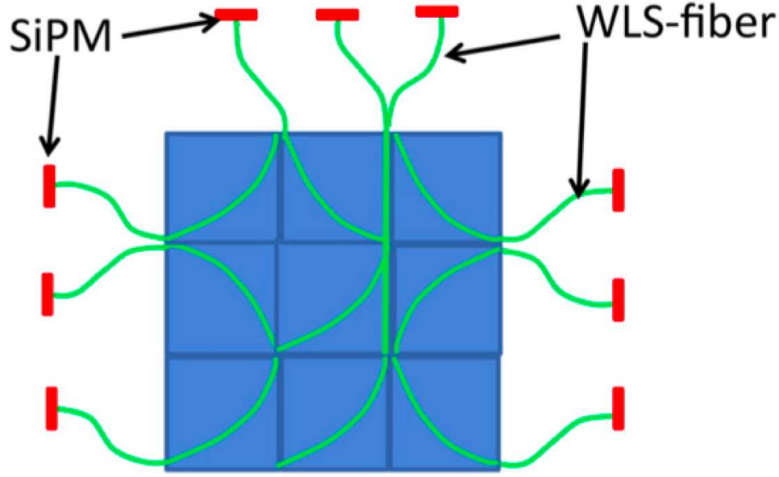}
\vspace{0.cm}{\caption{Schematic illustration of a scintillator-fiber-tungsten module with a lateral $3\times 3$ cells  segmentation. Wavelength shifting fibers pass laterally  through the plates and are read out at the side of the module  with either SiPMs or APDs photodetector.\label{moduleseg}}}
\end{center}
\end{figure}

To improve the $e/\pi$ rejection factor the calorimeter design can be done more granular with the fine longitudinal and transverse segmentation. One possible option of transverse calorimeter segmentation is shown in Fig.~\ref{moduleseg}. Here, instead of uniform scintillator plate, nine smaller size  scintillator tiles are inserted in each active layer of the calorimeter. The  lateral size $3\times 3 $ cm$^2$ of each smaller tile approximately corresponds to the Moliere radius that reliably identifies  the transverse profile of the e-m shower. The geometrical arrangement of the WLS-fibers  allows the light readout of each smaller scintillator tile by the photodetectors placed at three lateral sides of the calorimeter module as shown in Fig.~\ref{moduleseg}. To reduce the number of the readout channels the WLS-fibers from each three subsequent tiles can be grouped into one bunch viewed by a single photodetector. This option of the light readout leaves the calorimeter longitudinal segmentation fine enough with the three radiation length in each of ten sections. Taking into account the transverse segmentation, the total number of the readout channels per one module is equal to 90. Obviously, such dense  light readout configuration requires compact and inexpensive photodetectors. The silicon photomultipliers, SiPMs seam to be a natural candidates  in our case due to their compactness, relatively low cost, high gain and high photon detection efficiency. At present, there are  few types of SiPMs with high pixel density and, respectively, with high dynamical range, acceptable for the calorimetry. A few companies, such as Hamamatsu, KETEK and Zecotec,  can provide such photodetectors  with the required parameters.

We also studied the Moliere radius of this design. The fraction of the energy of a shower contained within a given radius (in terms of radiation length) for a calorimeter with one radiation length sampling and 3 mm scintillator was simulated. For pure tungsten, the Moliere radius is $R_M\simeq 2.6 ~X_0\simeq 9.3$ mm, and is the radius that contains approximately 90\% of the shower energy. From the simplified simulation, we can see that in order to absorb  nearly 90\% of the energy in the counter, its lateral size should be still  within roughly one $R_M$. It was also found that this value is nearly independent of energy from 1-40 GeV. The Moliere radius of this configuration is almost the same as that of pure tungsten, it is larger by about 20\%.

To estimate the $e/\pi$ rejection factor, we have performed also simulation of the ECAL1 response to the hadrons. A good overall $e/\pi$ suppression factor $\lesssim 10^{-3}$ could be expected based on detail description of the electromagnetic and hadronic shower profiles, both lateral and longitudinal, and their fluctuations in the calorimeters. An example of a developed technique  allowing accurate description of the electromagnetic shower shapes can be found in Ref.~\cite{Petti:2000xa,Autiero:1998uu}. The results of this work are planed to be used for further development of the method, including event-by-event shower shape fluctuations. An example of work related to the description of the lateral fluctuations of the hadronic showers can be found in Ref.~\cite{Gninenko:1998zu, Autiero:1997ya}.  

\begin{figure}
\begin{center}
\includegraphics[width=.5\textwidth]{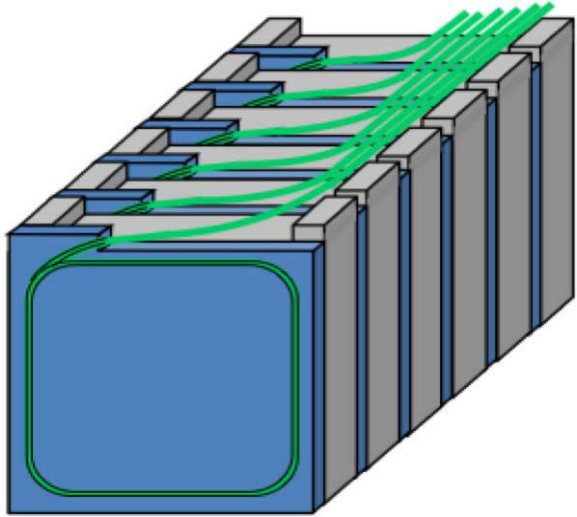}
\caption{Schematic illustration of scintillator-fiber-lead HCAL  module consisting of a stack of lead  and scintillator plates of the tzhickness 16 mm and 4 mm, respectively. Wavelength shifting fibers pass laterally  through the plates and are read out at the side of the module  with MAPD photodetectors.\label{module}}
\end{center}
\end{figure}

\subsubsection{Hadronic calorimeter} \label{sec:HadrCal}

\begin{figure}[tbh!]
\begin{center}
\includegraphics[width=.5\textwidth]{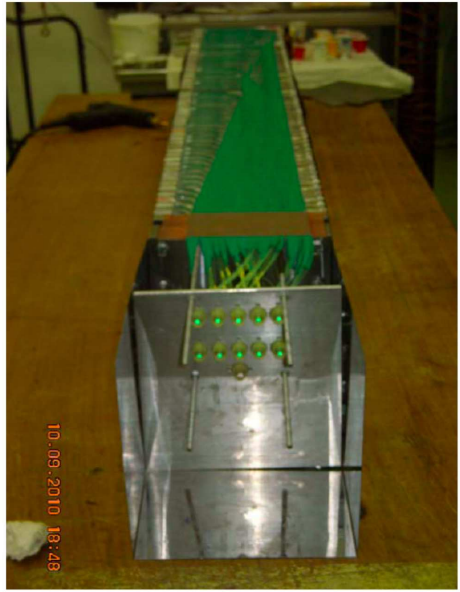}
\caption{Photo of the  HCAL module.\label{photo}}
\end{center}
\end{figure}

The hadronic calorimeter (HCAL) shown in Fig.~\ref{setup}, is used to enhance the longitudinal setup hermeticity for the sensitive search for the invisible decay $A' \rightarrow \chi \bar{\chi}$ into the lighter dark matter particles $\chi$, see Sec.~\ref{sec:ExpInvisible}. The HCAL consists of 4 modules~\cite{Golubeva:2009zza:Ivashkin:2012fd}. Each module consists of 60 lead/scintillator layers with 16 mm and 4 mm thickness, respectively, see Figs.~\ref{module} and~\ref{photo}. The lead/scintillator plates are tied together with 0.5 mm thick steel tape and placed in a box made of 0.5 mm thick steel. Steel tape and box are spot-welded together providing appropriate mechanical rigidity. The full length of modules corresponds to 5.7 nuclear interaction lengths. The module has  transverse dimension of $20\times 20$ cm$^2$ and weight 500 kg. The mechanical rigidity of these heavy modules was enhanced by a slight modification of their structure. Namely, one 16 mm lead layer in the middle of the module was replaced by a steel plate with similar nuclear interaction length. Light read-out is provided by Kyraray Y11 WLS-fibers embedded in round grooves in the scintillator plates. The WLS-fibers from each 6 consecutive scintillator tiles are collected together in a single optical connector at the end of the module. Each of the 10 optical connectors at the downstream face of the module is read-out by a single photo-diode. The longitudinal segmentation into 10 sections ensures good uniformity of light collection along the module and delivers information on the type of particle which caused the observed particle shower. 10 photodetectors per module are placed at the rear side of the module together with the front-end-electronics. The dependence of obtained energy resolution on beam energy in the energy range 10 - 200 GeV is given by $\frac{\sigma_E}{E} = \frac{0.57}{\sqrt{E}}+0.037$.

\subsubsection{Readout of Scintillating Fibers}

For the readout of the fibers and tiles it is planned to use the well-established waveform digitizing technology used already in several experiments, see e.g.\ the MEG experiment at PSI. This technology is based on the switched capacitor array chip DRS4 developed at PSI, which is capable of sampling the SiPM signal with up to 5 Giga samples per second with a resolution close to 12 bits. The advantage of this technology compared to traditional constant fraction discriminators and TDCs is that pile-up can be effectively recognized and corrected for. In addition, pulse height information becomes available which can be used to discriminate signals.

\subsubsection{The decay volume}

The decay volume is a tank of 30 cm in diameter and a length in the range from 3 m to 5 m. The volume is evacuated to the pressure below $10^{-3}$ mbar, to minimize secondary hadronic interactions in air. The in- and output flanges are made of a thin Mylar layers, about 20 mg/cm$^2$. 
  
\begin{table}[tbh!]
\caption{\bf Design, purpose, performance, and event rate of the detectors used in the experiment.\label{setupsum}}  
\begin{tabular}{lr}
\hline
\hline
{\bf 1. Electromagnetic  calorimeters ECAL1 and ECAL2}  \\
$\bullet$ design:  sandwich  (3.5 mm W + 3 mm Sc) $\times$ 30 layers \\
$\bullet$ purpose: energy measurements, shower profile measurements, $e/\pi$ separation \\
$\bullet$ performance: energy resolution $\Delta E/E \simeq 0.18/\sqrt{E}$, X,Y resolution $\simeq 3$ mm, \\
~~~$e/\pi$ rejection $\lesssim10^{-2}$ \\
$\bullet$ event rate: up to $10^7$ $e^-$ per spill, $10^{12}-10^{13}$ $e^-$ in total run\\
\hline
{\bf 2. Hadronic Calorimeter HCAL}  \\
$\bullet$ design: sandwich (16 mm Pb + 4 mm Sc) $\times$ 60 layers \\
$\bullet$ purpose: $\pi, p, n$ detection  \\
$\bullet$ performance: energy resolution $\Delta E/E \simeq 0.55/\sqrt{E}$, $\pi$-hermeticity $\simeq10^{-8}$ \\
$\bullet$ event rate: up to $10^5$ $\pi$ per spill, $10^{10}-10^{11}$ in total run\\
\hline
{\bf 3. Beam counters S1 and S2  }  \\
$\bullet$ design: Sc 1mm fiber hodoscopes \\
$\bullet$ purpose: $e^- e^+$ pair hits and track detection   \\
$\bullet$ performance: spacial  resolution $ \simeq 1 $ mm, 2 tracks separation  $\Delta R \gtrsim 1$ mm \\
$\bullet$ event rate: up to $10^5$ $e^-$ per spill\\
\hline
{\bf 4. Veto counters }  \\
$\bullet$ design: plastic scintillator \\
$\bullet$ purpose: low energy charged track detection\\
$\bullet$ performance: mip inefficiency  $\lesssim 10^{-.4}$ \\
$\bullet$ event rate:  up to $10^5$ hits per spill\\
\hline
{\bf 5. Synchrotron photon counter } \\
$\bullet$ design: 5 mm thick LYSO crystal \\
$\bullet$ purpose: X-ray energy mesurements \\
$\bullet$ performance: energy resolution $\Delta E/E \simeq 30\%$ at 50 keV, time resolution $\simeq 1$ ns. \\
$\bullet$ event rate: up to $10^7$ 10-100 keV $\g$ per spill, $10^{12}-10^{13}$ for full run\\
\hline
{\bf 6. Decay volume}  \\
$\bullet$ design: diameter $\simeq$30 cm $\times$ 5 m length, filled with He or vacuum $\lesssim 10^{-5}$ Torr\\
$\bullet$ purpose:  minimize secondary particles interactions \\
\hline 
\hline
\end{tabular}
\end{table}

\subsubsection{Data taking and trigger}\label{sec:DAQ}

To define a valid electron event hitting the calorimeter we have requested that the beam counters (not shown in Fig.~\ref{setup}) are in coincidence. This condition defines the  beam Trigger~1. The total area covered by the  beam is $\simeq 10 \times 10$ mm$^2$. When counters S1 and S2 are added to the coincidence, defining Trigger 1 condition, we get the so called Trigger 2. This trigger starts DAQ. The estimated events rate is well below 1 kHz. 

In Table~\ref{setupsum} the design, purpose, performance, and event rate of the detectors used in the experiment are summarized. The description of the counter to detect synchrotron radiation photons (item 5 of Table) is given below in Sec.~\ref{sec:ExpInvisible}.

\subsection{Background}
 
The background processes for the $\aee$ decay signature $S_{A'}$ of \eqref{sign} can be  due to physical-  and  beam-related sources. To perform full  simulation of the setup in order to investigate these backgrounds down to the level  $ \lesssim 10^{-12}$  would require a very large number of generated events resulting in a prohibitively large amount of computation time. Consequently, only the following, identified as the most dangerous background processes are considered and evaluated  with  reasonable statistics combined  with numerical calculations:

\subsubsection{\texorpdfstring{$\g,e^-$}{gamma, e-} - punchthrough}

\begin{itemize} 
\item The leak of the primary electron energy into the ECAL2, could be due to the  bremsstrahlung process $e^- Z  \rightarrow e^- Z \gamma$, when the emitted photon carries away almost all initial energy, while the final state electron with the much lower energy $E_{e^-}\simeq 0.1 E_0$ is absorbed in the ECAL1. The bremsstrahlung photon could punch through the ECAL1 and V1  without interactions, and  produce an $\ee$ pair  in the S1, which deposit  all its energy in the ECAL2. The photon could also be absorbed in a photonuclear reaction occurring in the ECAL1 and resulting in, e.g.\ an energetic leading secondary neutron. 
  
In the first case, to suppress this background, one has to use the ECAL1 of enough thickness, and as low  veto energy threshold as possible. Assuming that the primary interaction vertex  is selected to be within a few first $X_0$s, for the total remaining ECAL1+V1 thickness of $\simeq 30$ $X_0$, the probability for a photon to punch through both ECAL1 and V1 without interaction is $\lesssim 10^{-13}$. Thus, this background is at the negligible level. In the second case, an estimation results in a similar background level $\lesssim 10^{-13}$.

\item Punch-through primary electrons, which penetrate the ECAL1 and V1 without depositing much energy could produce a fake signal event. It is found that this is also an extremely rare event.

\end{itemize}

The  beam-related background can be due to a  beam particle misidentified as an electron. This background is caused by some pion, proton and muon contamination in the electron beam.  

\subsubsection{Hadronic background }

\begin{itemize} 
 
\item The first source of this type of background could be due to the  
\begin{equation}
p (\pi) + A \rightarrow n + \pi^0 + X,~n \rightarrow \mathrm{ECAL2} \label{prob}
\end{equation}
reaction chain: i) an incident hadron produces a neutral pion with the energy $E_{\pi^0} \lesssim 0.1  E_0$ and an energetic leading neutral hadron, e.g.\ neutron, carrying the rest of the energy of the primary collision  with the nucleus (A,Z), ii) the neutral pion decays $\pi^0 \rightarrow 2\g$ generating an e-m shower in the ECAL1, while iii) the neutron penetrates the rest of the ECAL1 and the veto counter V1  without interactions, scatters in the counter S1, producing low energy secondaries and deposits all its energy in the ECAL2. The probability for such a reaction chain to occur can be estimated as 
\begin{equation}
P_{p (\pi)} \simeq f_{p (\pi)} \cdot P_{\pi^0n} \cdot P_{S1} \cdot P_{n} \, ,
\end{equation}
where $f_{p(\pi)}, ~P_{\pi^0n}, ~P_{S1}, ~P_{n}$ are, respectively, the level of the admixture of hadrons in the primary beam, $P_{p(\pi)} \lesssim  10^{-2}$, the probability for an incoming hadron to produce the $\pi^0 n $ pair in the ECAL1,  $P_{\pi^0n} \simeq 10^{-4}$, the probability for the neutron to interact in S1, $P_{S1}\simeq 10^{-3}$, and the  probability for the leading $n$ to deposit all its energy in the ECAL2, $P_{n}\simeq 10^{-3}$. This results in $P\lesssim 10^{-12}$. The probability for neutral hadrons to interact in the S1 of thickness $\simeq $ 1 mm, or $\simeq 10^{-3}$ nuclear interaction length,  can be reduced significantly, down to $P_{S1}\simeq 10^{-4}$, by replacing it, e.g.\ with a  wire chamber counter. This leads to $P\lesssim 10^{-13}$. At low energies $E_0\lesssim 30$ GeV, the requirement to have two hits in the S1 would  suppress the background further.

Note, that the cross section for the reaction $p (\pi)  + A \rightarrow \pi^0 + n + X  $, with the leading neutron in the final state, has not yet been studied in detail for the wide class of nuclei and full range of hadron energies. To perform an estimate of the $P_{\pi^0 n}$ value, we use data from the ISR experiment at CERN, which studied leading neutron production in $pp$ collisions at $\sqrt{s}$ in the range from 20 to 60~GeV~\cite{Flauger:1976ju,Engler:1974nz}. For these energies the invariant cross sections, obtained as a function of $x_F$ (Feynman $x$) and $p_T$, were found to be in the range $0.1  \lesssim  E\frac{d^3\sigma}{d^3p} \lesssim 10$ mb/GeV$^2$ for $0.9 \lesssim x_F\lesssim  1$ and  $0 \lesssim p_T\lesssim 0.6$ GeV~\cite{Flauger:1976ju}. Taking these results into account, the cross sections for leading neutron production in our energy range are estimated  by using the Bourquin-Gaillard formula, which gives the parametrized  form of the invariant cross section for the production in high-energy hadronic collisions of different hadrons over the full phase-space, for more details  see,  e.g.~\cite{Gninenko:2012eq}. The leading neutron  production cross sections in $p(\pi)A$ collisions are evaluated  from its linear extrapolation to the target atomic number.
  
In another case, the leading neutron could interact in a very downstream part of the veto counter producing leading $\pi^0$  without being detected. The $\pi^0$ decays  subsequently into $2\g$ or $\ee \g$. The background from this events chain is also estimated  to be very small.

\item The fake signature $S_{A'}$ arises when the incoming pion produces in a very upstream part of the ECAL1 a low energy neutral pion, escapes detection  in the V1 counter due to its inefficiency, and either deposits all its energy in the ECAL2, or decays in flight in the DV into an $e\nu$ pair with the subsequent decay electron energy deposition in the ECAL2. In the first case, also relevant to protons, an analysis similar to the previous one, shows that this background is expected to be at the level $\lesssim 10^{-13}$. In the second case, taking into account the probability for the $\pi \rightarrow e \nu $ decay in flight, and that the electron would typically have about one half of the pion energy, results in a suppression of this background to the level $< 10^{-15}$.   

\end{itemize}
The overall probability of the fake signal produced by an incoming hadron is estimated to be $P_{p (\pi)} \lesssim 10^{-13}$ per incoming electron. Another type of background is caused by the muon contamination in the beam.

\subsubsection{Muon background}
\begin{itemize}
\item  The muon could produce a low energy  bremsstrahlung photon in the ECAL1, which would be absorbed in the detector, then penetrates the V1 without being detected, and after producing signals in the S1 and S2 counters, deposit all its energy in the ECAL2 through the  emission of a hard  photon:  
\begin{equation}
\mu + Z \rightarrow  \gamma + \mu + Z, ~\mu \rightarrow {\rm ECAL2} \, .
\label{muon}
\end{equation}
The probability for the chain \eqref{muon} is estimated to be $P \lesssim 10^{-14}$. Similar to \eqref{prob}, this estimate is obtained assuming  that the muon contamination in the beam is $\lesssim 10^{-2}$, the probability for the muon to cross the V1 counter without being detected is  $\lesssim 10^{-4}$, and the probability for the $\mu$ to deposited all its energy in the ECAL2 is $\lesssim 10^{-7}$. Here, it is also taken into account that the muon should stop in the ECAL2 calorimeter completely to avoid being detected in the counter V2. An additional suppression factor arises from the requirement to have two-mip's  signal in the decay counters.

\item One more  background source can be due the event  chain 
\begin{equation}
\mu + Z \rightarrow \mu + \g + Z,~ \mu \rightarrow e \nu \nu,
\end{equation} 
when the incoming muon produces in the initial ECAL1 part a low energy bremsstrahlung photon, escapes detection  in the counter V1, and then decays in flight in the DV into $e\nu \nu$. There  are several suppression factors for this background: i) the relatively long muon lifetime resulting in a small probability to decay, ii) the presence of two neutrinos in the $\mu$ decay. The energy deposition of decay electrons in the ECAL2 is typically significantly  smaller  than the primary energy $E_0$, and iii) the requirement to have double mip energy deposition in the beam counters S1 and S2. All these factors  lead to  the expectation for this  background level to be at least $\lesssim 10^{-14}$.

\item A random superposition of uncorrelated events during the detector gate time could also results in a fake signal. Taking into account the selection criteria of signal events results  in a the small number of these background events $\lesssim 10^{-14}$. 
\end{itemize}

The overall probability of the fake signal from  muons  is estimated to be $P_\mu \lesssim 10^{-14}$ per incoming electron, and the accidental background is below $\lesssim 10^{-14}$.

\begin{table} 
\begin{center}
\caption{Expected contributions from different background sources estimated for the beam energy 100 GeV (see text for details).\label{tablevis}}
\vspace{0.15cm}
\begin{tabular}{lr}
Source of background& Expected level\\
\hline
punchthrough $e^-$s or $\g$s& $ \lesssim 10^{-13}$\\
hadronic reactions & $ \lesssim 2\times  10^{-13}$\\
$\mu$ reactions  & $ \lesssim 10^{-14}$\\
accidentals  & $\lesssim 10^{-14}$\\
\hline 
Total (conservatively)  &         $ \lesssim  3 \times 10^{-13}$\\
\end{tabular}
\end{center}
\end{table}
In Table~\ref{tablevis} contributions from all background sources are summarized for the beam energy of 100 GeV. The dependence on the energy is rather weak. The total background level is conservatively  $\lesssim 3\cdot 10^{-13}$, and is dominated by the admixture of hadrons in the electron beam. Thus, a search accumulating up to $\simeq 10^{13}$ $e^-$ events, is expected to be background free. 

\subsubsection{Direct measurements of the background level.}
To evaluate the background in the signal region one could perform independent direct measurements of its level with  the same setup by using pion and muon beams of proper energies. For this purpose the primary beam is tuned to pions. The muons can be selected by putting thick absorber on the primary beam line.

\subsection{Sensitivity of the experiment} \label{sec:SensitivityVisible}
To estimate the sensitivity of the proposed experiment  a simplified feasibility study based on GEANT4~\cite{Agostinelli:2002hh:Allison:2006ve} Monte Carlo simulations has been performed for 30 and 150 GeV electrons. The energy threshold in the ECAL1 is taken to be 0.5 GeV. The reported further analysis also  takes into account passive materials from  the walls of the decay vessel.

The significance of the $\aee$ decay  discovery  with the described  detector scales as~\cite{Bityukov:1998ju,Bityukov:2004kp} 
\begin{equation}
S=2\cdot(\sqrt{n_{A'} + n_b}-\sqrt{n_b}) \, ,
\label{sens}
\end{equation}
where  $n_{A'}$ is the number of observed signal events (or the upper limit of the observed number of events), and $n_b $ is the number of  background events. 

For a given number of electrons  on the target of length $L'$, $n_{e}\cdot t$ (here, $n_e$ is the electron beam intensity and  $t$ is the experiment running time) and $A'$ flux $dn_{A'}/dE_{A'}$,  the expected number of $\xdecay$   decays occurring within the fiducial volume  of the DV with the subsequent energy deposition in the ECAL2 calorimeter, located  at a distance $L$ from the $A'$ production vertex is given by 
\begin{equation}
n_{A'} \sim n_e t \int  A \frac{d n_{A'}}{dE_{A'}} exp\bigl(-\frac{L'M_{A'}}{p_{A'}\tau_A'}\bigr) \bigl[1-exp\bigl(-\frac{L M_{A'}}{p_{A'}\tau_A'}\bigr)\bigr]  \frac{\Gamma_{\ee}}{\Gamma_{tot}}  \varepsilon_{\ee}  dE_{A'}dV \, ,
\label{nev}
\end{equation}
where $p_{A'}$ is the $A'$ momentum, $\tau_{A'}$ is the $A'$ lifetime at the rest frame, $\Gamma_{\pair},~\Gamma_{tot}$ are the  partial and total $A'$-decay widths, respectively, and  $\varepsilon_{\ee}(\simeq 0.9)$ is the $\pair$ pair reconstruction efficiency. The flux of $A'$s produced in the process \eqref{reaction} is calculated by using the  $A'$ production cross section in the $e^- Z$ collisions from Ref.~\cite{Bjorken:2009mm}. The acceptance $A$ of the ECAL2 calorimeter is calculated  tracing $A'$s produced in the ECAL1 to the ECAL2, and is close to 100\% (see Section~\ref{sec:Theory}).
\begin{figure}[tbh!]
\begin{center}
\includegraphics[width=0.7\textwidth]{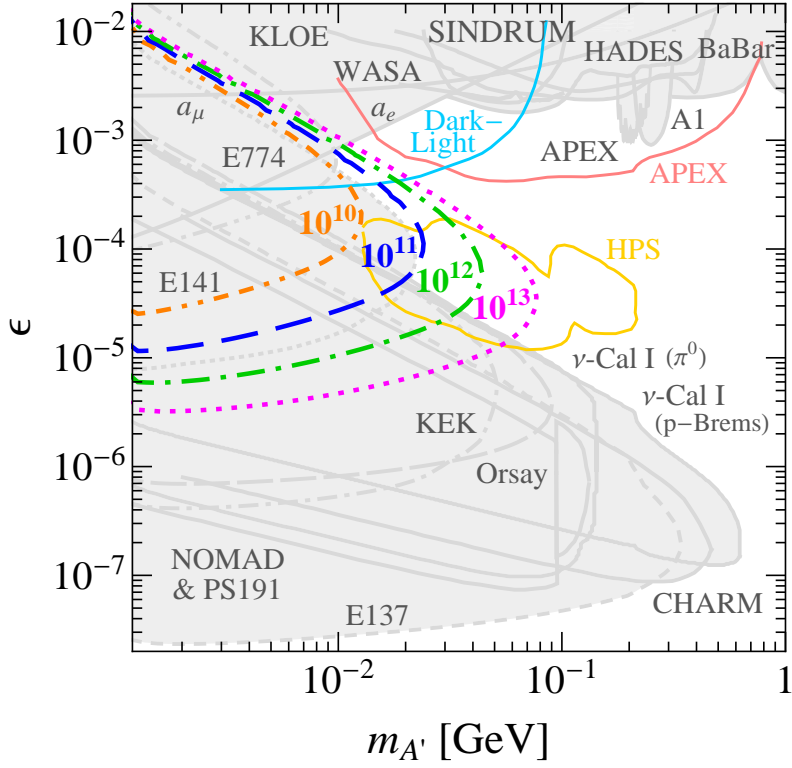}
\caption{Expected 90\% C.L.\ exclusion areas  in the ($M_{A'}; \epsilon$) plane for the collected data statistics of $ 10^{10},~ 10^{11}, ~10^{12}, ~10^{13}$ $e^-$ at 30 GeV. Shown are in gray all areas which are currently excluded by different other searches, see text for details. Expected sensitivities of the planned  APEX (full run),  DarkLight and HPS experiments  are  also shown for comparison~\cite{Hewett:2012ns}. For a review of all experiments, which  intend   to probe a similar parameter space, see Ref.~\cite{Hewett:2012ns, Essig:2013lka} and references therein.\label{plot}}
\end{center}
\end{figure}

If no excess events are found, the obtained results can be used to impose bounds on the $\gamma-A'$ mixing strength as a function of the  dark photon mass. Taking Eqs.~\eqref{rate}, \eqref{sens} and~\eqref{nev} and into account  and using the relation $ n_{A'}(M_{A'}) < n_{A'}^{90\%}(M_{A'}) $, where $n_{A'}^{90\%}(M_{A'})$  is the 90\% C.L.\ upper limit for the  number of signal events from the decays of the $A'$ with a  given mass $M_{A'}$  one can  determine the expected $90\%$ C.L.\ exclusion area in the ($M_{A'}; \epsilon $) plane from the results of the experiment. For the  background free case \textit{($ n_{A'}^{90\%}(M_{A'}) = 2.3$~events)}, the exclusion regions corresponding to accumulated statistics $10^{11}, ~10^{12},~10^{13}$ $e^-$ at 30 GeV (H4-30) are shown in Fig.~\ref{plot}. One can see,  that these exclusion areas are  complementary to the ones expected from the planned  APEX (full run), HPS   and DarkLight experiments, which   are  also shown for comparison~\cite{Hewett:2012ns, Essig:2013lka}. For a review of all experiments, which  intend   to probe a similar parameter space, see Ref.~\cite{Hewett:2012ns, Essig:2013lka} and references therein. Shown are also areas excluded from the electron (g-2) considerations ($a_e$ and $a_\mu$)~\cite{Endo:2012hp,Pospelov:2008zw}, by the results of the electron beam-dump experiments E141~\cite{Bjorken:2009mm,Riordan:1987aw}, E137~\cite{Bjorken:2009mm,Bjorken:1988as}, E774~\cite{Bjorken:2009mm,Bross:1989mp}, KEK~\cite{Andreas:2012mt,Konaka:1986cb} and LAL Orsay~\cite{Andreas:2012mt,Davier:1989wz}, the electron thin target experiments A1 at MAMI~\cite{Merkel:2011ze} and APEX~\cite{Abrahamyan:2011gv}, cf.\ also~\cite{Beranek:2013nqa}, by the $\nu$-Cal I experiment~\cite{Blumlein:2011mv,Blumlein:2013cua}, by the KLOE collaboration~\cite{Archilli:2011zc}, by data of the experiment SINDRUM~\cite{Gninenko:2013sr,MeijerDrees:1992kd}, by the WASA-at-COSY collaboration~\cite{Adlarson:2013eza}, by the HADES collaboration~\cite{Agakishiev:2013fwl}.

The statistical limit on the sensitivity of the proposed experiment is proportional to $\epsilon^2$. Thus, it is important to accumulate a large number of events. As one can see from Eq.~\eqref{nev}, the obtained exclusion regions  are also sensitive to the choice of the length $L'$ of the calorimeter ECAL1, which should be as short as possible. As discussed in Sec.~\ref{sec:SPSbeamline}, assuming the maximal secondary  H4 beam rate $n_e \simeq 5 \times 10^6~e^-$/spill at $E_0 \simeq 30-50$ GeV, we anticipate $\simeq 3 \times 10^{12}$ collected $e^-$s during $\simeq$ 3 months of running time for the experiment. Note, that since the decay time of the scintillating-fiber light signal is $\tau \lesssim 50$ ns, the maximally allowed electron  counting rate in order to avoid significant pileup effect is, roughly $ \lesssim 1/ \tau \simeq 10^{7}~ e^-$/s. This is well compatible with the maximal beam rate during the 4.8 s spill, which is expected to be $\lesssim 10^7/4.8 s \simeq 2 \times 10^6$. To minimize dead time, one could use a first-level trigger rejecting events with the ECAL2 energy deposition less than, say, the energy $\simeq 0.9 E_{0}$ and, hence, run the experiment at a higher event rate.

In the case of the signal observation, to cross-check the result, one could  remove the decay vessel DV and put the calorimeter ECAL2 behind the ECAL1. This would not affect the main background sources and  still allow the $A'$s production, but  with their decays upstream  of the ECAL2 calorimeter being suppressed. The distributions of the energy deposition in the ECAL1 and ECAL2 in this case would contain mainly background events, while  the signal level from the decays $\aee$ should  be reduced. The background  can also be independently studied with the muon and pion beams of the same energy. The evaluation of the $A'$ mass value could be  obtained from the results of measurements at different distances $L$ and beam energies. Finally note, that the performed analysis for the sensitivity of the proposed experiment may be strengthened  by more accurate and  detailed simulations of the H4 beam line and concrete experimental setup.\vspace{0.3cm}


\section{The experiment to search for the decay \texorpdfstring{$\ainv$}{A' -> invisible}} \label{sec:ExpInvisible}

The $A'$s could  also decay invisibly into a pair of dark matter particles $\chi \bar{\chi}$, see~\cite{Essig:2013lka, Essig:2013vha} and references therein. 

\begin{figure}[tbh!]
\begin{center}
\includegraphics[width=0.6\textwidth]{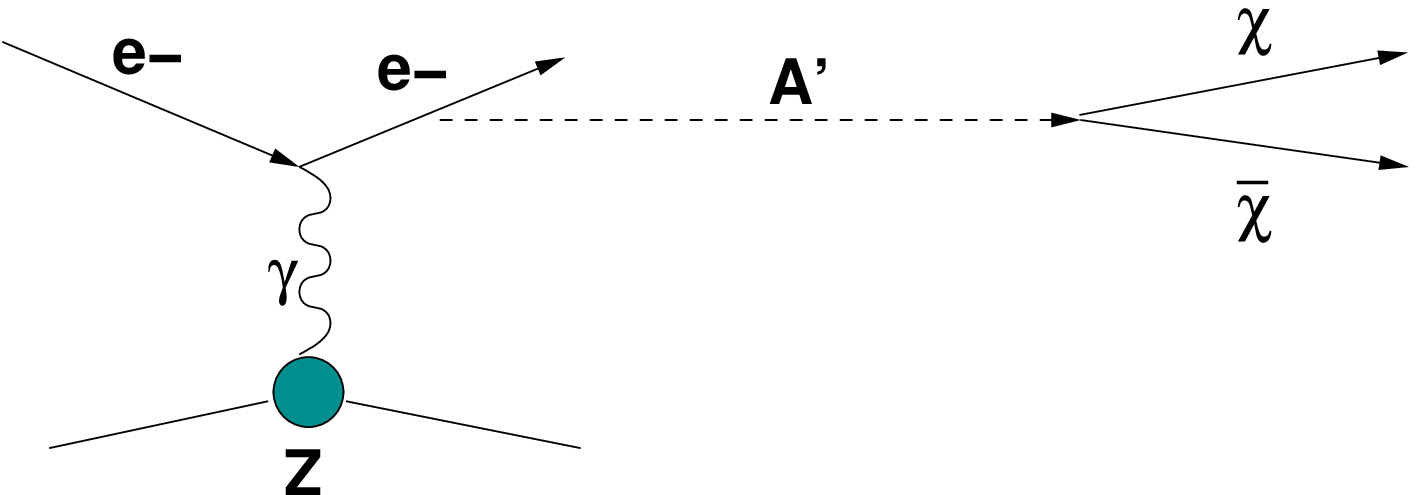}
\caption{Diagram illustrating the massive $A'$ production in the  reaction $e^- Z \rightarrow e^- Z A'$ of electrons scattering off a nuclei (A,Z) with the subsequent $A'$ decay into a $\inv $ pair.\label{diagrinv}}
\end{center}
\end{figure}
The diagram for the $A'$ production in the reaction 
\begin{equation}
 e^- Z \rightarrow e^- Z A',~\ainv 
\label{reactioninv}
\end{equation}
is shown in Fig.~\ref{diagrinv}.

The process of the dark photon production and subsequent invisible decay $\ainv$ is also expected to be  a very rare  event. For the previously mentioned parameter space, it is  expected to occur with the rate \textit{$\lesssim 10^{-10}$} with respect to the ordinary photon production rate. Hence, its observation presents a challenge for the detector design and performance.

\subsection{The setup}
The experimental setup specifically designed to search for the $A'\rightarrow invisible$ decays is schematically shown in Fig.~\ref{setup}. The experiment  employs the same very clean high energy $e^-$ beam. The admixture of the other charged particles in the beam  (beam purity) is below $10^{-2}$. The detector shown in Fig.~\ref{setup} is equipped with a  high density, compact electromagnetic (e-m) calorimeter ECAL1 to detect $e^-$ primary interactions, high efficiency veto counters V1 and V2, two scintillating  fiber counters (or proportional chambers) S1, S2 and a combination of the electromagnetic  calorimeter ECAL2 and HCAL located at the downstream end of the $A'$ decay volume DV to detect all final state products from the primary reaction $e^- Z \rightarrow e^- Z A'$. 

The method of the search is the following. The $A'$s are produced through the mixing with bremsstrahlung photons from the electrons scattering off nuclei in the ECAL1. The reaction~\eqref{reactioninv} typically occurs in the first few radiation length ($X_0$) of the detector. The bremsstrahlung $A'$ then either penetrates the rest of the setup  without interactions  and decays in flight into an $\ee$ pair outside the detector, or it could decay invisibly, $\ainv$, into two dark matter particles which also penetrate the rest of the setup without interaction. Similar to the previous case, the  fraction $f$ of the primary beam energy $E_1 = f E_0$  is deposited in the ECAL1. The ECAL1's downstream part is served  as a dump to absorb completely the e-m shower tail. For the radiation length $\lesssim$ 1 cm, and the total thickness of the ECAL1 $\simeq 30~X_0$ (rad.\ lengths) the energy leak  from the ECAL1 into the V1 is  negligibly small. The remained part of the primary electron energy $E_2 = (1-f)E_0$ is either transmitted trough the rest of the ECAL1 and other detector  by the $A'$, or is carried away by the products of the decay $\ainv$. In order to suppress  background due to inefficiency of detection (see below),  the detector must be longitudinally completely hermetic. To enhance detector hermeticity, a hadronic calorimeter (HCAL) with a total thickness  $\simeq 15-20  ~\lambda_{int}$ is placed behind the ECAL2, as shown in Fig.~\ref{setup}. If we assume that the $A'$ decays dominantly into the invisible final state, then the calorimeter ECAL1 is not constrained in length anymore, as it was in the case of $\aee$ decays. Then, the ECAL1(and ECAL2) calorimeter could be, e.g.\ a hodoscope  array of  the lead tungstate (PWO) heavy crystal counters ($X_0 \simeq 0.89$ cm), each of the size $10\times 10 \times 300$ mm$^3$, allowing accurate measurements of the lateral and longitudinal shower shape. The energy resolution of such calorimeters is quite good. As a function of the beam energy it is given by  $\frac{\sigma}{E} = \frac{2.8 \%}{\sqrt{E}} \oplus 0.4\% \oplus \frac{142~MeV}{E}$  \cite{Adzic:2009aa}. 
 
\begin{figure}[tbh!]
\begin{center}
\includegraphics[width=0.7\textwidth]{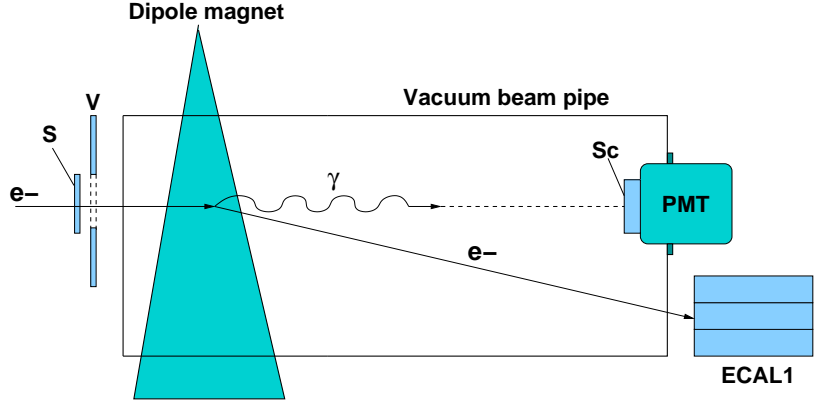}
\caption{The scheme of the additional tagging of high energy electrons  in the beam by using the electron synchrotron radiation in the banding magnetic dipole. The synchrotron radiation photons are detected by a $\gamma$ - detector by using the LYSO inorganic crystal (Sc) capable for the work in vacuum. The crystal is viewed by a high quantum efficiency photodetector, e.g.\ PMT, SiPM, or APD. The beam defining counters S and veto V are also shown.\label{tag}}
\end{center}
\end{figure}

The occurrence of $\ainv$ decays produced in $e^- Z $ interactions would appear as an excess of events with a single e-m showers in the ECAL1, Fig.~\ref{setup}, and zero energy deposition in the rest of the detector, above those expected from the background sources. The signal candidate events have the signature:  
\begin{equation}
S_{A'} = {\rm ECAL1 \times \overline{V1 \times S1 \times S2 \times ECAL2 \times V2 \times HCAL}},
\label{signinv}
\end{equation}
and should satisfy the following selection criteria:  
\begin{itemize}
\item The starting point of (e-m) showers in the ECAL1 should be localized  within a few first $X_0$s.  
\item The lateral and longitudinal shapes of the shower in the ECAL1 are consistent with an electromagnetic one. The fraction of the total  energy deposition in the ECAL1 is $f\lesssim 0.1$, while in the ECAL2 it is zero.
\item No energy deposition in  the V1,V2, ECAL2, and HCAL.
\end{itemize}
To improve the primary high energy electrons selection  and additionally suppress background from the possible presence of low energy electrons in the beam typically with energy $E_e \lesssim 0.1 E_0$ (see below), one can use a high energy $e^-$-tagging system utilizing  the synchrotron radiation (SR) from high energy electrons in a dipole magnet, as schematically shown in Fig.~\ref{tag}. The basic idea is that, since the critical SR photon energy is $(\hbar \omega)^c_\g \propto E_0^3$ (here $E_0$ is the beam energy) the low energy electrons in the beam could be rejected by  using, e.g.\ the cut on $E_\g> 0.3 (\hbar \omega)^c_\g$ in a X-ray detector, shown, for example in Fig.~\ref{tag}. In this scheme, the  electrons and radiation photons are detected separately. The total length of the vacuum line is about 100 m. The possibility of identifying electrons by detecting their synchrotron radiation with the xenon filled multi-wire proportional chamber, has been demonstrated previously, see e.g.~\cite{Dworkin:1986tk}. Preliminary, we consider the scheme shown in Fig.~\ref{tag} for detection of the synchrotron radiation photons in vacuum by utilizing inorganic LYSO crystal with a high light yield. Note that  electrons with the  energy $\lesssim 10$ GeV which are present in the beam before the dipole magnet will be deflected by it at a large angle, so they do not hit the ECAL1. However, such electrons could appear in the  beam after the magnet due to the muon $\mu \rightarrow e \nu \nu$ or pion $\pi \rightarrow e \nu $ decays in flight in the vacuum beam pipe. Since $\mu$s and $\pi$s do not radiate in the magnet, this source of the background is supposed to be suppressed.

\subsection{Background}
 
The background processes for the $\ainv$ decay signature $S_{A'}$ of~\eqref{signinv} can be classified as being due to physical- and beam-related sources. They could be due to the calorimeters energy resolution, cracks and beam holes in the setup. Unfortunately, direct measurements of the background level for the $\ainv$ decay mode is practically impossible, because of unknown low-energy tail in the beam electron energy distribution. So, our main goal in the detector design was not to try to reduce any background source to its possible lowest level, but only below the physical background. Similar to the decay channel $\aee$, we face familiar problems:  to perform full detector simulation in order to investigate these backgrounds down to the level  $ \lesssim 10^{-10}$  would require a huge number of generated events resulting in a prohibitively large amount of computer time. Consequently, only the following background sources, identified as the most dangerous processes are considered and evaluated  with  reasonable statistics combined  with numerical calculations. 

\subsubsection{Electron background}

\begin{itemize}
\item The leak of the primary electron energy, could be due to the  bremsstrahlung process $e^- Z  \rightarrow e^- Z \gamma$,  when the emitted  photon carries away almost all of its initial energy, while the final state electron with the much lower energy  $E_{e^-}\simeq 0.1 E_0$ is absorbed in the ECAL1. The photon could  punch through  the rest of the detector  without interactions. The photon could also be absorbed  in a photo-nuclear reaction occurred in the ECAL1 resulting in, e.g.\ an energetic leading secondary punch-through neutron. 

In the first case, to suppress this background, one has to use the  ECAL2+HCAL  of enough thickness,  and  a  low  veto threshold as possible. Taking into account that the primary interaction vertex  is selected to be within  a few first $X_0$s, for the total remaining ECAL1+V1 thickness of $\simeq 30$ $X_0$, the probability for a photon to punch through it without interaction is $\lesssim 10^{-13}$. Thus, this background is at the negligible level. In the second case, the analysis results in a similar background level $\lesssim 10^{-13}$.

\item Punch-through primary electrons, which penetrate the ECAL1  without depositing much energy could produce a fake signal event. It is found that this is also an extremely rate event.

\item Contributions due to the detector material non-uniformity, presence of cracks, and beam holes result effectively to a  degradation of the energy resolution, e.g.\ in the vicinity of a hole. However, due to the proper design of the HCAL the contribution from  these effects to the HCAL inefficiency is found to be negligible. The possible effect caused by the HCAL module support structure (stainless tapes of 1 mm thick) could be minimized by positioning the HCAL at a small angle with respect to the  beam axis.  
\begin{figure}[tbh!]
\begin{center}
\includegraphics[width=0.6\textwidth]{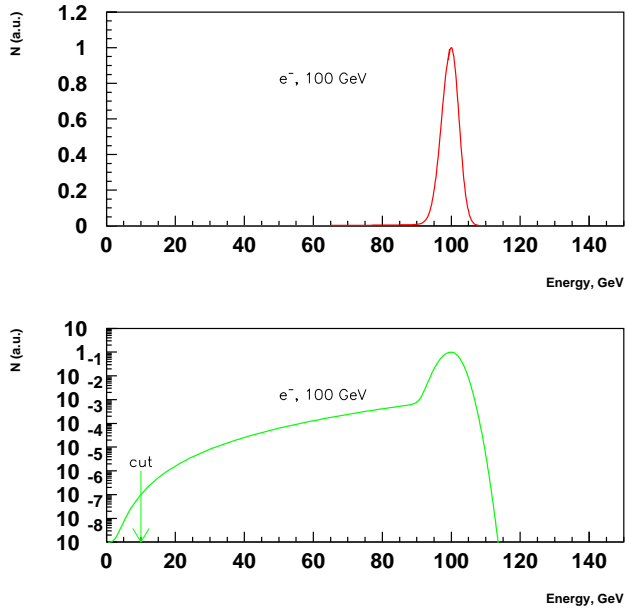}
\caption{Hypothetical energy distributions of 100 GeV electrons in a SPS secondary electron  beam, shown in normal- (top) and log-scales (bottom). The arrow shows the selection cut on the energy deposited in ECAL1, see Fig.~\ref{setup}. The fraction of events below the cut determines the sensitivity of the $\aee$ decay search.\label{tail}}
\end{center}
\end{figure}

\item One of the main background sources is related to the  low-energy tail in the electron energy distribution in the primary beam.  The electrons are selected and tuned to a given momentum by a few hundreds meters spectrometer. The origin of the low-energy tail is caused by the beam electron interactions with a passive material in the beam, such e.g.\ as entrance windows of the beam lines, residual gas, etc... Another source of the low energy tail is related to the pion or muon decays in flight in the beam line. To predict the fraction of events in the tail and their energy distribution is not simple. A full beam line simulation at a high level of precision has to be performed. Just for  illustrative purposes,  in Fig.~\ref{tail} the hypothetical e$^-$ spectra with a low energy tail are shown for a 100 GeV beam. The sensitivity of the experiment is determined by the fraction of electron events with energy $E$ below of a certain threshold $E_{th}$ : $\frac{N(E< E_{th})}{N_0}$.  For example, for the primary beam energy of 100 GeV and $E_{th}\simeq 10 GeV$, this ratio is expected to be very small, probably well below 10$^{-6}-10^{-8}$. To additionally suppress this level of background one can use the electron tagging system based on the detection of  the synchrotron radiation  from high energy electrons in a dipole magnet as schematically shown in Fig.~\ref{tag}.

\item  The reaction  
\begin{equation}
e + Z \rightarrow  \gamma + e + Z, ~ \gamma + A \rightarrow {\rm invisible}
\label{photon}
\end{equation}
may occur: an electron can emit a hard bremsstrahlung photon and deposit the rest of its energy in the ECAL1. The photon could induce a photo-nuclear reaction accompanied by the emission of a leading neutral particle(s), such as neutron. The neutron then could be undetected in the rest of the setup. Taking into account the above estimated non-hermeticity of the detector, the probability of the  reaction~\eqref{photon} is found to $\lesssim 10^{-12}$.   

This reaction could also occur in the residual gas of the beam pipe located after the magnet. To reduce this possible background as good as possible vacuum in the  pipe is required.  
  
\item Finally, the electroproduction of a neutrino pair    
\begin{equation}
e + Z \rightarrow  e + Z + \nu \overline{\nu},
\label{neutrinos}
\end{equation}
resulting in the invisible final state accompanied by the recoil electron  energy deposition in the ECAL1 can occur. The preliminary estimate shows that the ratio of the cross sections for the reaction \eqref{neutrinos} to the bremsstrahlung cross section is $\lesssim 10^{-13}$. More accurate calculations are in progress \cite{nvk}.   

\end{itemize}

\begin{figure}[tbh!]
\begin{center}
\includegraphics[width=0.7\textwidth]{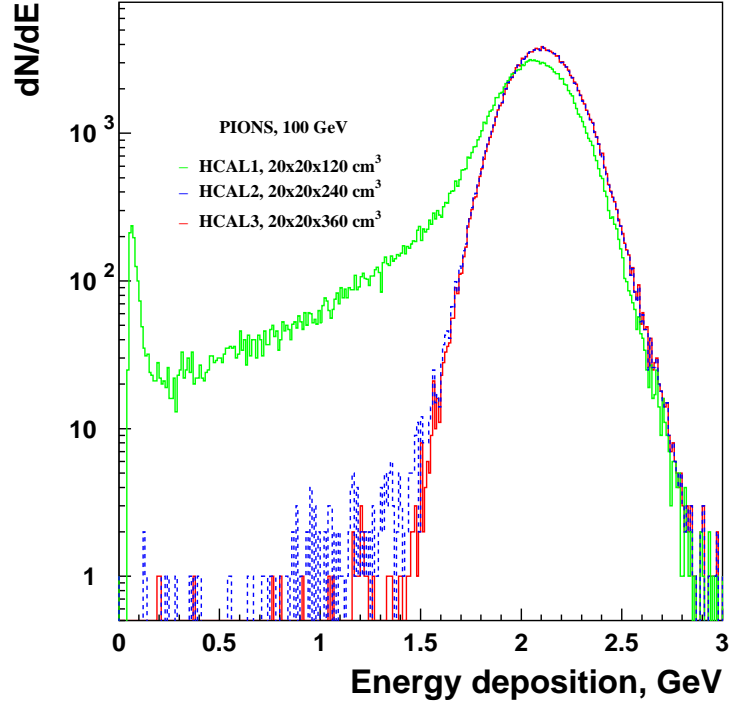}
\caption{Distributions of energy deposited by $2\times 10^5$ $\pi^-$ with energy 100 GeV in the three consecutive HCAL modules with the lateral size 20$\times$20 cm$^2$. The peak at 0.06 GeV corresponds to energy deposited by the punch-through pions.\label{pions}}
\end{center}
\end{figure}

\subsubsection{Hadronic  background }

The hadronic background can be, for example due to beam hadrons misidentified as  electrons. This background is caused by some pion, proton, etc.\ contamination in the electron beam. Another source of this type of background is caused by the hadron electroproduction in the ECAL1. 

For the measurement of the hadronic energy we used the following expression for the energy resolution of the HCAL 
\begin{equation}
\frac{\sigma (E)}{E} = \frac{0.55}{\sqrt{E}} + 0.037,  
\label{hcalres}  
\end{equation}
which corresponds to the case of the HCAL calorimeter constructed by the INR group for the experiment NA61, see Sec.~\ref{sec:HadrCal} and Ref.~\cite{Golubeva:2009zza:Ivashkin:2012fd}. The inefficiency for the zero energy detection, e.g.\ due to pile-up effects is estimated to be below 10\% assuming the intensity of $5\times 10^6~e^-$/spill.

\begin{itemize} 
\item The fake signature~\eqref{signinv} could arise when i) either a hadron from the beam produces in the very beginning of the ECAL1 a low energy neutral pion and escapes detection in the rest of the detector, or ii) an electroproduced hard hadron(s) $h$ from the reaction $eA \rightarrow e h X$ occurred in the very upstream part of the ECAL1 is not detected. In the first case, the background is supposed to be suppressed by the requirement of the  presence of the synchrotron photon in the beam line. The second source  requires a more detailed study. In this case the background could be caused by the incomplete longitudinal hermeticity of the  detector. That is, there might be a leak of energy due to production of leading neutral particles such as neutrons and/or $K^0_L$s, which penetrate the ECAL2 and HCAL without depositing energy above the certain threshold $E_{th}$. This is the energy cut on the sum of energy depositions in the ECAL2 and HCAL below which an event is considered as the ``zero-energy'' event in the ECAL2+HCAL. The punchthrough probability is defined by $exp(-\lambda_{int}L)$, and is of the order $10^{-9}$ for about 20 $\lambda_{int}$ thickness of the detector composed of two ECALs and three consecutive HCAL modules (here $\lambda_{int}$ is the nuclear interaction length). This value should be multiplied by a  conservative factor $\lesssim 10^{-4}$, which corresponds to the probability $P_{h}$ of a  single leading hadron production per incoming electron  in the ECAL1, resulting in the final value of $\lesssim 10^{-12}$.

\begin{figure}
\centering
\begin{minipage}{.5\textwidth}
  \centering
  \includegraphics[width=1.\linewidth]{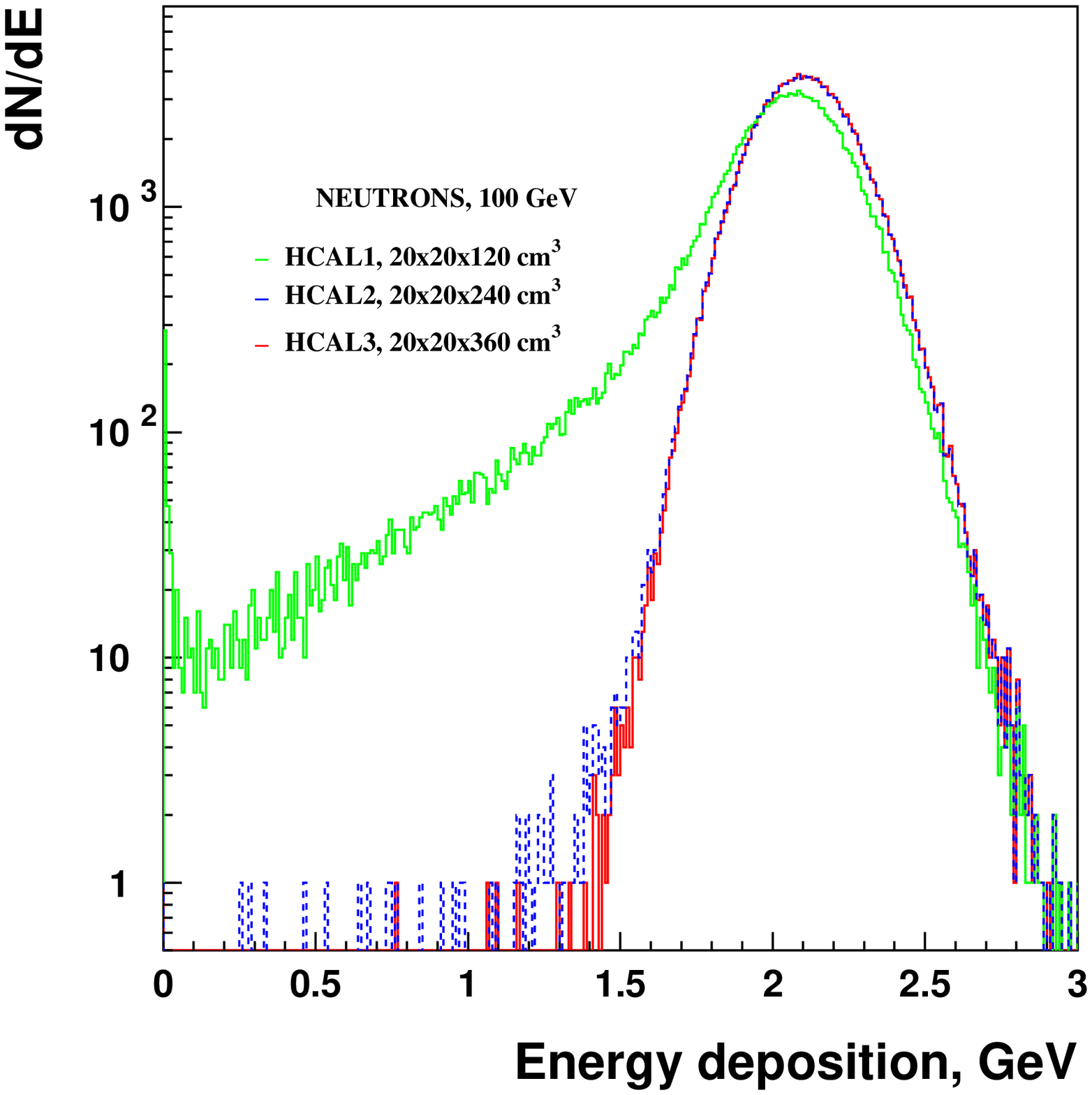}
\end{minipage}%
\begin{minipage}{.5\textwidth}
  \centering
  \includegraphics[width=1.\linewidth]{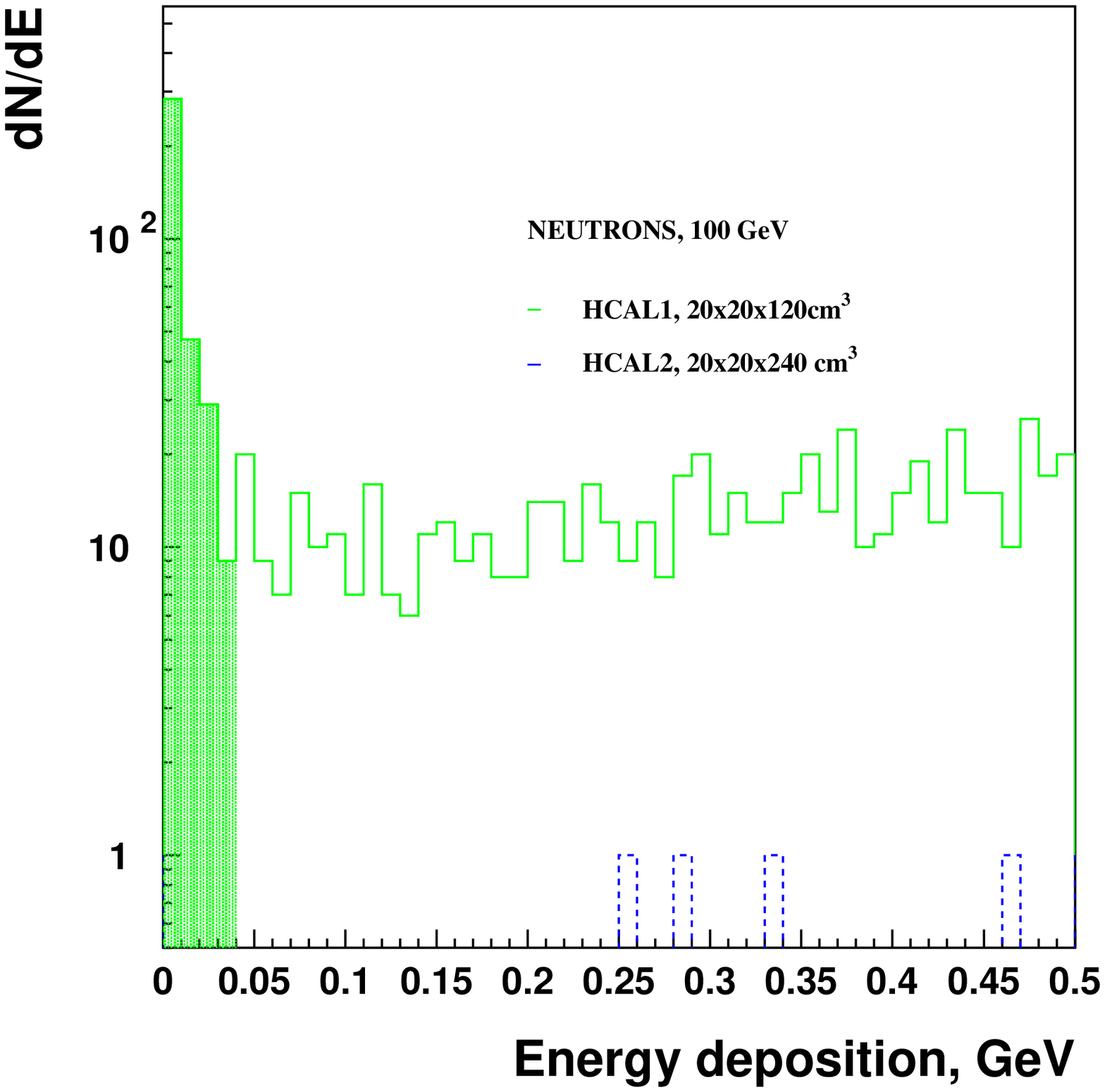}
\end{minipage}
\caption{Distributions of energy deposited by $2\times 10^5$ neutrons with energy 100 GeV in the three consecutive HCAL modules with the lateral size 20$\times$20 cm$^2$ (left side). Shown on the right is the low energy part of the spectrum. The peak at zero-energy (dashed) is due to the punch-through neutrons.\label{neutrons}}
\end{figure}

The HCAL non-hermeticity for high energy hadrons was estimated with a GEANT4-based simulation. In Fig.~\ref{pions} the simulated distributions of energy deposited by $2\times 10^5$  negative pions with energy 100 GeV in three consecutive HCAL modules with the lateral size 20$\times$20 cm$^2$ are shown. The peak at 0.06 GeV for the single module (HCAL1) corresponds to the punchthrough pions penetrating the HCAL without interaction. The fraction of events in the peak agrees well with the estimate of the punchthrough probability discussed  above. As expected, the peak disappears for the larger HCAL thickness. It can be noticed that for charged pions the HCAL is completely hermetic, i.e.\ there is always energy released by the pion in the detector. 

This picture is different for neutral hadrons. As an example, in Fig.~\ref{neutrons} the simulated distributions of energy deposited by $2\times 10^5$ neutrons with energy 100 GeV in three consecutive HCAL modules with the lateral size 20$\times$20 cm$^2$ are shown. One can see that for the single module case, there are events with zero-energy deposition in the HCAL1. These events  correspond to punchthrough  neutrons penetrating the detector without interaction. 

To estimate the HCAL non-hermeticity for higher neutron statistics an attempt was made to reduce the computational time by considering  only events from the tail of the deposited energy distribution. This tail is shown in Fig.~\ref{htail} for about $5\times10^6$ neutrons interacting in  the ECAL2 plus three consecutive HCAL modules assembly (see Fig.~\ref{setup}) at 100 GeV. 
\begin{figure}[tbh!]
\begin{center}
\includegraphics[width=0.7\textwidth]{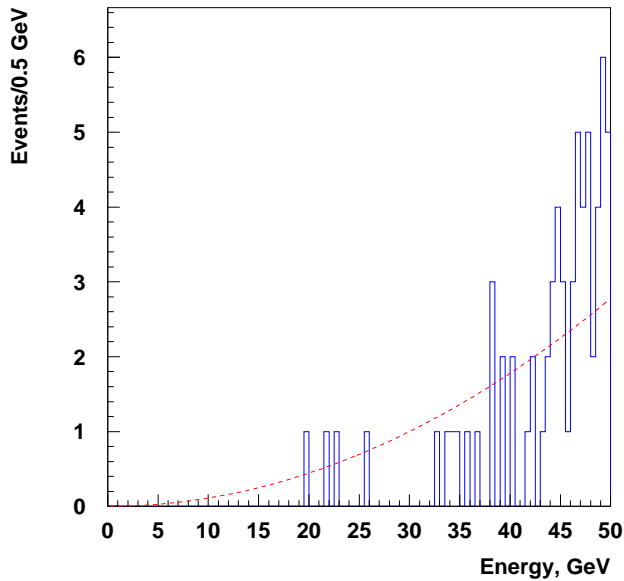}
\caption{The expected low energy tail distributions of sum of energies in the ECAL2+HCAL3 from about $5\times10^6$ neutrons with primary energy of 100 GeV, see Fig.~\ref{setup}. The peak from the total energy deposition in normalized to 100 GeV. The dashed curve shows the polynomial fit  to the distribution. The presence of possible cracks, holes, non-uniformuties is ignored.\label{htail}} 
\end{center}
\end{figure}
The obtained low energy tail distribution was fitted by a polynomial function, shown as red dashed line in Fig.~\ref{htail}, and the results were extrapolated to the lower energy part of the spectrum in order to evaluate non-hermeticity of the ECAL2+HCAL3 assembly at low $E_{th}$-values. In the experiment the threshold is expected to be around $E_{th}\simeq 1$  GeV. This procedure results in a (ECAL2+HCAL3)-non-hermeticity, defined as the ratio of the number of events below the threshold $E_{th}$ to the total number of incoming pions $\eta = n(E<E_{th})/n_{tot}$, which is shown in Fig.~\ref{nonherm}. One can see, that for the energy threshold $E_{th} \simeq 1$ GeV the non-hermeticity is expected to be at the level $\eta \lesssim 10^{-9}$.  Finally, taking into account the probability for the single leading hadron electro-production to be $P_{h} \lesssim 10^{-4}$, results in an overall level of this background of $\lesssim10^{-12}$.   
\begin{figure}[tbh!]
\begin{center}
\includegraphics[width=0.7\textwidth]{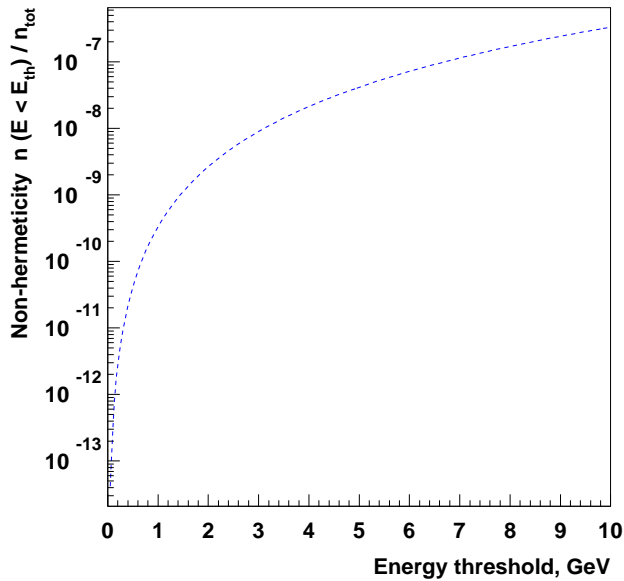}
\caption{The estimated non-hermeticity of the ECAL2+HCAL3 as a function of the energy threshold $E_{th}$ on the sum of energy deposition in both calorimeters ECAL2+HCAL3 for 100 GeV neutrons interacting in the assembly.\label{nonherm}}
\end{center}
\end{figure}

\end{itemize}
The overall probability of the fake signal produced by incoming pions or protons is conservatively estimated to be $P_{p (\pi)} \lesssim 10^{-12}$ per incoming electron. It should be noted, that in order to take data simultaneously for both, visible and invisible, $A'$ decay modes, the HCAL3  located downstream  the ECAL1,  has to be increased in lateral size in order to avoid background from large transverse hadronic shower fluctuations. The use of the HCAL3 detector with the cross section 40$\times$40 cm$^2$ (2$\times$2 HCAL modules) is foreseen in this case.   

Another type of background originates by the muon contamination in the beam.

\subsubsection{Muon background}

\begin{itemize}
\item  The muon could produce a low energy  photon  in the ECAL1, which would be  absorbed in the calorimeter and then penetrates the rest of the detector without being detected
\begin{equation}
\mu + Z \rightarrow  \gamma + \mu + Z, ~\mu \rightarrow {\rm invisible} \, .
\label{muoninv}
\end{equation}
The probability for the events chain~\eqref{muoninv} is estimated  to be  $P \lesssim 10^{-12}$. This estimate is obtained assuming  that the muon contamination in the beam is $\lesssim 10^{-2}$, the probability for the muon to cross the V1 and V2 without being detected is at least  $\lesssim 10^{-6}-10^{-8}$,  and the probability for the  $\mu$ to deposited its energy in the ECAL2+HCAL below the threshold $E_{th}$ is $\lesssim 10^{-4}$.

\item One more  background can be due the event chain 
\begin{equation}
\mu + Z \rightarrow \mu + \g + Z,~ \mu \rightarrow e \nu \nu,
\end{equation} 
when the incident  muon  produces in the initial ECAL1 part a low energy bremsstrahlung photon, escapes detection  in the V1, and then decays in flight in the DV into $e\nu \nu$, and the electron is non detected because its energy is either $E< E_{th}$ or due to pure HCAL energy resolution, if it misses the ECAL2. There  are several suppression factors for this  source of background: i) the relatively long muon lifetime resulting in a small probability to decay, ii) the presence of two neutrinos in the $\mu$ decay reduces the electron energy. However, it is practically not possible for the decay electron to avoid energy deposition in the ECAL2, because the electron is emitted at a small angle $\lesssim 50 {\rm MeV}/E_e$, and the probability for $E_e$ to be  significantly  smaller  than the primary energy $E_0$ is very low.  These factors  lead to  the expectation for this  background to be  at the level at least $\lesssim 10^{-12}$
\end{itemize}
The overall probability of the fake invisible signature \eqref{signinv} produced by  muons  is estimated to be $P_\mu \lesssim 10^{-12}$ per incoming electron. In Table~\ref{tab:table2} contributions from the all background processes are summarized for the beam energy of 100 GeV. The total background is conservatively at the level $\lesssim  10^{-12}$, and is dominated by the admixture of hadrons in the electron beam. This means that the search accumulated up to $\simeq 10^{12}$ $e^-$ events, is expected to be background free.     
\begin{table}[tbh!] 
\begin{center}
\caption{Expected contributions to the total level of background from different background sources estimated for the beam energy 100 GeV (see text for details).}\label{tab:table2}
\vspace{0.15cm}
\begin{tabular}{lr}
Source of background& Expected level\\
\hline
punchthrough $e^-$s or $\g$s& $ \lesssim 10^{-13}$\\
HCAL non-hermeticity & $ \lesssim   10^{-12}$\\
$e^-$'s  low energy tail, $E_e\lesssim 0.1 E_0$& $ \lesssim  10^{-12}$\\
$\mu$ reactions  & $ \lesssim 10^{-12}$\\
$e^-$ induced photo-nuclear reactions & $\lesssim 10^{-12}$\\
\hline 
Total (conservatively)  &         $ \lesssim   10^{-12}$\\
\end{tabular}
\end{center}
\end{table}

\subsection{Sensitivity of the experiment}

\begin{figure}[tbh!]
\begin{center}
\includegraphics[width=0.7\textwidth]{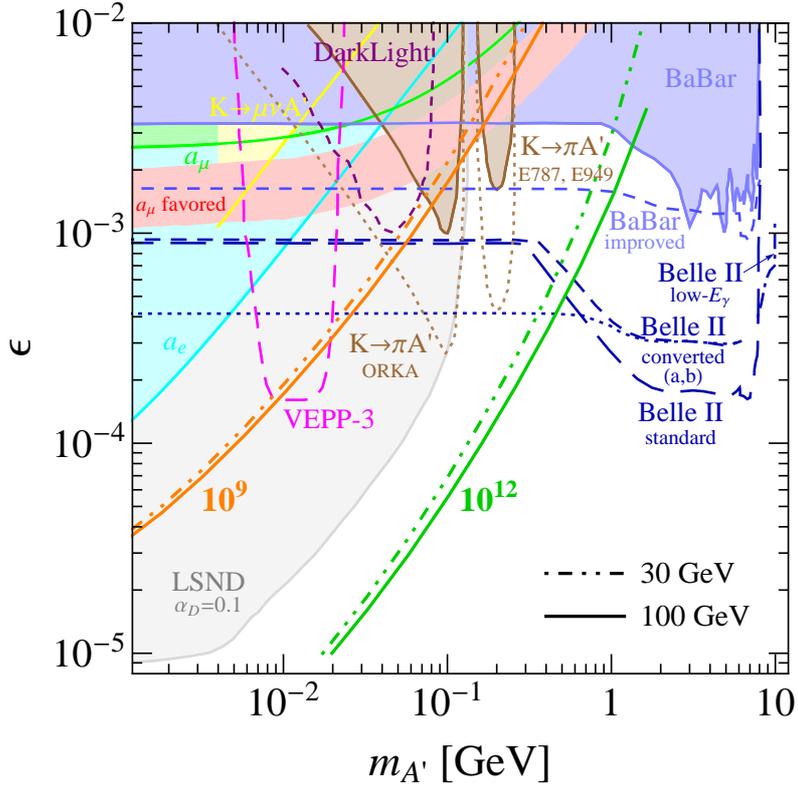}
\caption{Constraints in the $\epsilon$ vs $M_{A'}$ plane for invisibly decaying $A'$ assuming they can decay invisibly to a pair of dark-sector states $\chi\bar{\chi}$, provided  $M_{A'} > 2 m_\chi$. The orange and green lines show the expected 90\% C.L.\ exclusion areas corresponding, respectively, to $10^9$ and $10^{12}$ accumulated electrons at 30 GeV (dash-dotted) and 100 GeV (solid) for the background free case. Various other constraints (shaded regions) and projected sensitivities (dashed lines) are also shown, mostly adapted from Ref.~\cite{Essig:2013vha}. The constraint for the BaBar mono-photon search is given as blue shaded region, while the blue dashed lines represent the reach of an improved BaBar and possible Belle II mono-photon searches. Further limits are shown from the anomalous magnetic moment of the electron ($a_e$, cyan) and muon ($a_\mu$, green), the rare kaon decay $K^+ \rightarrow \pi^+ A'$ (brown) and leptonic decay $K^+ \rightarrow \mu^+ \nu_\mu A'$ (yellow)~\cite{Beranek:2012ey}, and LSND (light gray; assuming $\alpha_D$ = 0.1 and that $\chi$ can not decay to other light dark-sector states which do not interact with $A'$s)~\cite{deNiverville:2011it}. Other sensitivities are shown for the upcoming electron fixed-target experiments DarkLight (purple; shown when kinematically relevant) and VEPP-3 (magenta) as well as an improved sensitivity for $K^+ \rightarrow \pi^+ A'$ with ORKA (brown). The red shaded region is preferred in order to explain the discrepancy between the measured and the predicted SM value of $(g-2)_\mu$.\label{inv}}
\end{center}
\end{figure}

Using considerations, which are similar to those of Sec.~\ref{sec:SensitivityVisible}, the expected  exclusion areas in the plane $(\epsilon, M_{A'})$, shown in Fig.~\ref{inv}, are derived.  These areas  are shown for the background free case and correspond to accumulated statistics of $10^{9}$ (red line)  and $10^{12}$ (blue line)  $e^-$s with energy 100 GeV. The only assumption used is that the $A'$s decay dominantly to the invisible final state $\chi\bar{\chi}$,  if the $A'$ mass  $M_{A'} > 2 m_\chi$. One can see, that the area corresponding to  $10^{12}$ electrons completely covers  the LSND exclusion region obtained under the assumption of a certain $\chi-A'$ coupling strength $\alpha_D$. In Fig.~\ref{inv}, various other constraints are plotted as shaded regions and projected sensitivities of other experiments are indicated as lines. As suggested in~\cite{Izaguirre:2013uxa}, electron beam dump experiments searching for the light dark matter particle $\chi$ are sensitive to a similar region of the parameter space depending on $m_\chi$ and $\alpha_D$. Further limits on the invisible $A'$ have been derived from the cooling of white dwarfs for $m_\chi$ in the keV-range~\cite{Dreiner:2013tja} and from energy losses in supernova for $m_\chi$ in the MeV-range~\cite{Dreiner:2013mua}, both again assuming certain $\alpha_D$s. Constraints on dark matter particles charged under a hidden gauge group from primordial black holes~\cite{Dai:2009hx} do not apply to the mass range considered here. 

Similar to the case of the visible $A'$ decay search, the statistical limit on the sensitivity of the proposed experiment to search for decay channels $\ainv$,  is proportional to $\epsilon^2$ and is set mostly by its value and by  possible background. Thus, it is important to accumulate a large number of events. As discussed in Sec.~\ref{sec:SPSbeamline} we anticipate up to $\simeq 3 \times 10^{12}$ collected $e^-$s during $\simeq$ 3 months of running time for the experiment. In the case of the $\ainv$ signal observation,  several methods could be used to cross-check the result. For instance, one could perform measurements taken with different HCAL thicknesses. If the fake signal is due to the HCAL non-hermeticity, its expected level  can be obtained by extrapolating the results to a very large (infinite) HCAL thickness.

\section {Conclusion} \label{sec:Conclusions}

Due to their specific properties, dark photons are  an interesting probe of well motivated physics beyond the standard model both from the theoretical and  experimental point of view. We  propose to perform a  light-shining-through-a-wall experiment dedicated to the sensitive search for dark photons  in the still unexplored area of the mixing strength  $10^{-5}\lesssim \epsilon \lesssim 10^{-3}$ and masses $M_{A'} \lesssim 100$ MeV by using available 10-300 GeV electron beams from the CERN SPS. If  $A'$s exist, their di-electron decays $\aee$ could be observed by looking for events with the two-shower  topology of energy deposition in the detector. The key point for the experiment is an observation of events with almost all beam energy deposition in the ECAL2 calorimeter, located behind the ``ECAL1 wall''. Since the $A'$s are short-lived particles, the sensitivity of the search is $\propto \epsilon^2$, differently from the case of a search for  a long-lived $A'$, where the number of signal events is $\propto \epsilon^4$. 

In this proposal, we show that the sensitivity  of the  search for the $\aee$ decay in  ratio of cross sections $\frac{\sigma(e^-Z\rightarrow e^-Z A')}{\sigma(e^-Z \rightarrow e^- Z \g)}$ at the level of $\lesssim 10^{-13}-10^{-12}$ could be achieved. This sensitivity can be obtained with a setup  optimized for several of its properties: i) the intensity and purity of the primary electron beam, ii) the high efficiency of the veto counters, iii) a high number of photoelectrons from decays counters S1 and S2,  iv) the good energy and time resolution as well as capability to measure accurately longitudinal and lateral shape of showers in both ECAL1 and ECAL2 calorimeters. Large amount of collected electrons and background suppression are crucial to improve the sensitivity of the search. To obtain the best sensitivity for a particular parameters region,  the choice of the energy and intensity  of the beam, as well as the background level should be compromised. In the case of non-observation, the expected  exclusion areas are  complementary to the ones from the planned  APEX (full run), DarkLight, and other  experiments intended to probe a similar parameter space~\cite{Hewett:2012ns, Essig:2013lka}. 

The experiment has also the capability for a sensitive search for $A'$s decaying invisibly to dark-sector particles,  such as dark matter. Our feasibility study shows that a sensitivity for the search of the $\ainv$ decay mode in branching fraction $Br(A') = \frac{\sigma(e^-Z\rightarrow e^-Z A'), \ainv}{\sigma(e^-Z \rightarrow e^- Z \g)}$ at the level below a few parts in $10^{11}-10^{12}$ is in reach. The intrinsic background due to the presence of low energy electrons in the beam can be  suppressed by using a tagging system, which is based on the detection of synchrotron radiation of high energy electrons. The search would allow  to cover a significant fraction of the yet unexplored  parameters space for the $\ainv$ decay mode.

After testing the detector, that might commence in 2014-2015, the experiment would be performed in two phases. In the first phase in 2015, the goal is to optimize the detector components and measure the dominant backgrounds from  the hadron (and possibly muon) contaminations in the electron beam. This could be done  by using any secondary beam line of the SPS that would provide enough intensity in the given energy range for the background measurements. In the second phase, 2015-2016,  the goal is to reach the previously mentioned  sensitivity  or better by exploiting a possible  upgrade of the detector, which might be necessary given the results of phase I. To reach this goal utilizing  a secondary SPS beam line that would provide enough electron intensity for the signal search is mandatory. If an excess consistent with the signal hypothesis is observed, this would unambiguously indicate the presence of new physics. The full running time of the proposed measurements is requested to be up to several months, and it could be taken at different SPS secondary beams. Due to the moderated cost of the experiment, the required resources in terms of man power, equipment and consumables would already be available.

\section*{Acknowledgments}

We would like to thank L.~Di Lella, D.~S.~Gorbunov, and A.~A.~Radionov for useful discussions and A.~Fabich for valuable comments on the CERN SPS beam lines. SA acknowledges the support of the ERC project 267117 (DARK) hosted by Universit\'{e} Pierre et Marie Curie - Paris 6, PI J.~Silk.

\end{document}